\documentclass[5p,twocolumn,times]{elsarticle}

\usepackage{natbib}
\biboptions{sort&compress}
\usepackage{graphicx}
\usepackage{tikz}
\usepackage{tabularx}
\usepackage{babel}[english]
\usetikzlibrary{shapes,arrows}
\usepackage{array}
\usetikzlibrary{shapes.geometric}
\newcolumntype{L}[1]{>{\raggedright\let\newline\\\arraybackslash\hspace{0pt}}m{#1}}
\newcolumntype{C}[1]{>{\centering\let\newline\\\arraybackslash\hspace{0pt}}m{#1}}
\newcolumntype{R}[1]{>{\raggedleft\let\newline\\\arraybackslash\hspace{0pt}}m{#1}}
\usepackage{enumerate}
\usepackage{dcolumn}
\usepackage{xcolor}
\usepackage{bigints}
\usepackage{dblfloatfix}
\usepackage{upgreek}

\usepackage{comment}

\usepackage[colorlinks=true,urlcolor=blue]{hyperref} 

\usepackage{xspace}
\usepackage{booktabs}
\usepackage{multirow}
\usepackage{adjustbox}
\usepackage{relsize}
\usepackage{colortbl}
\usepackage{steinmetz}
\usepackage{scalerel}
\usepackage{float}
\usepackage{subcaption}
\usepackage{caption}

\usepackage{lineno}

\usepackage[per-mode=symbol]{siunitx}

\journal{Computers and Fluids}

\newcommand*{\Strouhal}{\textrm{St}}
\newcommand*{\T}{\mathsf{T}}

\usepackage{nomencl}
\usepackage{mdframed} 

\makenomenclature
\usepackage{etoolbox}
\renewcommand\nomgroup[1]{%
  \item[\bfseries
  \ifstrequal{#1}{M}{Mathematical Operators}{%
  \ifstrequal{#1}{A}{Abbreviations}{%
  \ifstrequal{#1}{R}{Roman Symbols}{%
  \ifstrequal{#1}{G}{Greek Symbols}{%
  \ifstrequal{#1}{S}{Subscripts and Superscripts}{%
  \ifstrequal{#1}{N}{Dimensionless Numbers}{}}}}}}%
]}

\begin{document}

\begin{frontmatter}

    \title{Meshless data-driven decompositions with RBF-based inner products}

    \author[1,2]{\texorpdfstring{M. Ratz\corref{cor1}}{M. Ratz}}
    
    \ead{manuel.ratz@vki.ac.be}
     \author[2,4,5]{A. Parente}
    \author[1,2,3]{M. A. Mendez}
    \cortext[cor1]{Corresponding author}
    \address[1]{von Karman Institute for Fluid Dynamics, Waterloosesteenweg 72, Sint-Genesius-Rode, Belgium}
    \address[2]{Aero-Thermo-Mechanics Laboratory, École Polytechnique de Bruxelles, Université Libre de Bruxelles, Av. Franklin Roosevelt 50, Brussels, 1050, Belgium}
    \address[3]{Experimental Aerodynamics and Propulsion Lab, Universidad Carlos III de Madrid, Av. de la Universidad 30, 28911 Leganés, Spain}
    \address[4]{Brussels Institute for Thermal-Fluid Systems and Clean Energy (BRITE), Brussels, 1050, Belgium}
    \address[5]{WEL Research Institute, 1300 Wavre, Belgium}
  
      \date{\today}
    
    \begin{abstract}
    Data-driven modal decompositions are useful tools for compressing data or identifying dominant structures. Popular ones like the dynamic mode decomposition (DMD) and the proper orthogonal decomposition (POD) are defined with continuous inner products. These are usually approximated with samples of data uniform in space and time. However, not every dataset fulfills this requirement. Numerical simulations with smoothed particle hydrodynamics or experiments with Lagrangian particle tracking velocimetry produce scattered data varying in time and space, rendering sample-based inner products impossible. In this work, we extend a previous approach that computes the modal decompositions with meshfree radial basis functions (RBFs). We regress the data and use the continuous representation of the RBFs to compute the required inner products. We choose our basis to be constant in time, greatly reducing the computational cost since the inner product of the data reduces to the inner product of the basis functions. We use this approach in the most popular decompositions, namely the POD, DMD, multi-scale POD, and the two versions of the spectral POD. For all decompositions, the RBFs give a mesh-free representation of the spatial structures. Two test cases are considered: particle image velocimetry measurements of an impinging jet and large eddy simulations of the flow past a transitional airfoil. In both cases, the RBF-based approach outperforms classical binning and better recovers relevant structures across all data densities.
    \end{abstract}

\end{frontmatter}



\section{Introduction}\label{sec:introduction}

Data-driven modal decompositions such as proper orthogonal decomposition (POD) and dynamic mode decomposition (DMD) are widely used to extract dominant structures and dynamics from complex systems. These techniques have been extensively applied to obtain compact representations of high-dimensional data across diverse fields, including fluid mechanics \cite{Berkooz1993,Rowley2017,Taira2017}, image processing and filtering \cite{Mendez2017}, face recognition \cite{Kirby1990}, reacting flows \cite{Procacci2022}, structural fault detection \cite{Fassois2007} and vibration \cite{Feeny1998}, biomechanics \cite{Troje2002}, neuroscience \cite{Nonnenmacher2017}, epidemiology \cite{Shi2014} to mention but a few.

Despite their wide applicability, the algorithms used to compute data-driven modal decompositions are traditionally formulated under the assumption that data are available on a spatial grid---as in image processing, experimental techniques such as particle image velocimetry (PIV, \cite{Raffel2018}), or standard numerical frameworks like finite difference, finite volume, and finite element methods \cite{LeVeque2007}. This grid may be structured or unstructured, dense or sparse, but is typically assumed to be fixed in time. Such an assumption simplifies computing the inner products inherent to any linear decomposition \cite{Mendez2023}, from classical methods like POD \citep{Lumley1967, Lumley1981, Sirovich1987} and DMD \citep{Rowley2017, Schmid2010} to more recent variants such as spectral POD (SPOD) \cite{Sieber2016, Towne2018} and multi-scale POD (mPOD) \cite{Mendez2019, Mendez2020}. The spatial structures on this discrete grid also allow to link dominant features to specific physical locations.

A variant within this context is the so-called gappy scenario, where an underlying grid or spatial structure is still assumed, but the available data have values missing in space at scattered locations in time. A common strategy in this setting is to resort to interpolation \cite{Stein1999}, radial basis functions \cite{Fasshauer2007}, or low-rank matrix completion \cite{Candes2009} to reconstruct the missing data either prior to or during the computation of a modal decomposition. Examples of such approaches include gappy POD \cite{Oliver1990,Everson1995,Gunes2006}, the recently proposed gappy Spectral POD \cite{Nekkanti2023}, and interpolative variants of DMD designed to handle non-uniform data in space \cite{Menon2020} or time \cite{Li2022,Smith2023}. Alternatively, when interpolation or regression methods become ineffective due to excessive gappiness, compressed sensing techniques \cite{Donoho2006,Brunton2015} can be employed to exploit sparsity and directly reconstruct the dominant flow structures from limited measurements, enabling full-field recovery from observations restricted to a sparse subset of grid points. These ideas have opened the path toward compressed sensing formulations of POD \cite{Matulis2024,Xiong2021} and DMD \cite{Brunton2015,Fathi2018}.

A more significant departure from the gappy framework arises when the data are not simply incomplete, but intrinsically \emph{scattered}, with no underlying spatial grid or fixed set of sampling locations. This situation arises in Lagrangian computational methods, such as smoothed particle hydrodynamics \cite{Monaghan1992} or vortex methods \cite{Leonard1980}, where the computational particles themselves move with the flow. Similarly, in experimental techniques like Lagrangian particle tracking velocimetry \cite{Ouellette2006, Schanz2016, Tan2020, Schroeder2023}, measurements are obtained at the positions of tracer particles advected by the fluid, leading to data sampled at locations varying in both space and time.

To the authors' knowledge, there is no established framework for applying modal decompositions directly to data in their scattered, Lagrangian form. The most common strategy is to circumvent the irregularity by mapping the data onto a fixed spatial grid---typically through interpolating, projecting, or binning techniques---and then proceeding with traditional grid-based decomposition methods \cite{Aguei1987,Schobesberger2021,Springel2010}. While this preprocessing step enables to use established tools like POD (e.g. \cite{Wu2023}) and DMD (e.g. \cite{Marshall2023}), it introduces interpolation errors and may obscure important features of the original scattered data, particularly in regions with sparse sampling or rapid deformation. More advanced binning strategies iterate between mapping the data to the grid and computing the POD and progressively improve the spatial modes \cite{CortinaFernandez2021, GrilleGuerra2024b}.

Within the image velocimetry community, significant effort has recently been devoted to methods for mapping scattered data onto fixed grids \cite{Gesemann2016, Jeon2022, Sciacchitano2025}. More recently, however, a meshless paradigm based on radial basis function (RBF) has begun to emerge \cite{Sperotto2022a, Sperotto2024a, Sperotto2024b, Li2024, Ratz2024, Rigutto2025}, aiming to remove grids entirely. These approaches are considered meshless since they operate directly on scattered data without requiring a structured or unstructured mesh, and do not assume a fixed spatial discretization for computing derivatives, integrals, or more complex tasks such as solving partial differential equations \cite{Fasshauer2007, Zhang2000, Chen2014, Fornberg2015}.

Recently, the meshless RBF framework was extended to compute a POD directly from scattered data \cite{Tirelli2025a}. The approach is conceptually related to functional principal component analysis (FPCA) \citep{ramsay2005principal, Wang2016, Hall2006}, which generalizes traditional principal component analysis (PCA) to the case where each observation is treated as a continuous function rather than a finite-dimensional vector. The key difference, among others discussed in \citet{Tirelli2025a}, is that the FPCA uses interpolation methods to approximate an infinite-dimensional eigenvalue problem, whereas the meshless POD retains the classical POD structure but replaces the grid-based inner product with RBF-based quadrature.

This work extends the framework introduced in \citet{Tirelli2025a} in several important respects. First, we replace the interpolative RBF formulation with a regression-based approach and substitute the Gauss–Legendre quadrature used to assemble the temporal correlation matrix with a fully RBF-based quadrature scheme. This modification not only reduces computational cost by enabling the reuse of precomputed weight matrices but also improves flexibility in handling complex geometries. Second, we introduce an RBF-based projection strategy operating directly on the continuous functional form obtained from the RBF approximation of the snapshots, enabling efficient evaluation of the spatial structure of each mode. Finally, the proposed framework is generalized beyond classical POD to encompass more advanced modal decompositions, including the two SPODs, mPOD, and DMD.

The remainder of the paper is organized as follows. Section~\ref{sec:2} recalls the fundamentals of RBF regression, how they are used to compute spatial inner products, and how each of the decompositions is expressed in the RBF framework. Section~\ref{sec:3} presents the two selected test cases and Section~\ref{sec:4} presents the results of the meshless algorithms compared to classic binning. Finally, Section~\ref{sec:5} closes with conclusions and perspectives.

\section{Mathematical framework}\label{sec:2}
The goal of all data-driven decompositions is to express data with respect to a set of basis functions. Here, we treat a two-dimensional velocity field in a domain $\Omega \in \mathbb{R}^2$, which we denote as $\boldsymbol{u}(\boldsymbol{x}, t) = \left({u}(\boldsymbol{x}, t),\,{v}(\boldsymbol{x}, t) \right)$, with $\boldsymbol{x} = (x, y)\in \Omega$ an (arbitrary) point in space and $t\in[0,T]$ an (arbitrary) point in time. The approach readily extends to higher dimensions and to scalar quantities such as temperature or density. 

The general modal decomposition, in a continuous sense, reads
\begin{align}
\label{deco}
    \boldsymbol{u}(\boldsymbol{x}, t) = \sum_r^{n_R} \sigma_r \boldsymbol{{\phi}}_r(\boldsymbol{x}) {\psi}_r(t)\,,  
\end{align} where $n_R$ is the number of modes, $\boldsymbol{{\phi}}_r(\boldsymbol{x})$ is a vector valued basis function for the spatial domain, and ${\psi}_r(t)$ is a scalar basis function for the time domain. These bases have unitary norm with respect to a certain measure in the spatial and temporal domain, such that the scalars $\sigma_r$ are amplitudes of the modal contribution. The inner product defining these norms and the associated projection operator are 
\begin{equation}
    \label{inner_T}
    \left\langle p(t), q(t) \right\rangle_T= \frac{1}{T}\int^T_0 q^\ast(t) p(t) dt 
\end{equation} for two continuous scalar functions $p(t),q(t)$ in $t\in[0,T]$ and 
\begin{equation}
    \label{inner_Omega}
    \left\langle \boldsymbol{p}(\boldsymbol{x}), \boldsymbol{q}(\boldsymbol{x}) \right\rangle_\Omega= \frac{1}{|\Omega|}\int_\Omega \boldsymbol{q}^{\ast}(\boldsymbol{x})\boldsymbol{p}(\boldsymbol{x}) d\boldsymbol{x}\,, 
\end{equation} for two continuous vector fields $\boldsymbol{p}(\boldsymbol{x}),\boldsymbol{q}(\boldsymbol{x})$ in $\boldsymbol{x}\in \Omega$, with $|\Omega|$ a measure of the domain (an area in 2D or a volume in 3D). Throughout the paper, we use the asterisk superscript to denote both a complex conjugate of a scalar or the Hermitian transpose of a vector or matrix.


All traditional grid-based decompositions use approximations of the integrals in \eqref{inner_T} and \eqref{inner_Omega} \citep{Mendez2023}. In the simplest case of uniform sampling in time, such that $\mathbf{t}=[\mathbf{t}_1, \mathbf{t}_2, \cdots , \mathbf{t}_{n_t} ]= \{\mathbf{t}_i\}^{n_t}_{i=1}$ is a vector collecting the equally distributed times $\mathbf{t}_i$ and $T=n_t \Delta t$, the samples of the functions $\textrm{p}(\mathbf{t}_i),\textrm{q}(\mathbf{t}_i)$ can be stored in vectors $\mathbf{p}, \mathbf{q}\in\mathbb{R}^{n_t}$ and the inner product product \eqref{inner_T} can be approximated as:
\begin{equation}
    \label{eq:inner_T_D}
    \langle p(t), q(t)\rangle_T\approx \langle {p}(\mathbf{t}_i),{q}(\mathbf{t}_i)\rangle_{T,d}= \frac{1}{n_t}\mathbf{q}^\ast \, \mathbf{p}\,,
\end{equation} with $\langle \cdot \rangle_{T,d}$ denoting the discrete approximation of the continuous inner product.

Similarly, in the spatial domain, denoting as $ \{ \mathbf{X}_n\}_{n=1}^{n_s}$ the set of $n_s$ sample locations in space, and assuming that each of these is associated to the same partitions of the domain $\Delta \Omega_i$ (an area in 2D or a volume in 3D), and denoting as $\mathbf{p},\mathbf{q}\in\mathbb{R}^{2n_s}$ the flattened matrix collecting the samples $\boldsymbol{p}(\mathbf{X}_i), \boldsymbol{q}(\mathbf{X}_i)$, \eqref{inner_Omega} can be approximated as 
\begin{equation}
    \label{eq:inner_Omega_D}
    \langle \boldsymbol{p}(\boldsymbol{x}), \boldsymbol{q}(\boldsymbol{x})\rangle_{\Omega}\approx \langle \boldsymbol{p}(\mathbf{X}_i),\boldsymbol{q}(\mathbf{X}_i)\rangle_{\Omega,d}= \frac{1}{n_s}\mathbf{q}^\ast \, \mathbf{p}\,,
\end{equation} having introduced the discrete spatial inner product. These discrete inner products \eqref{eq:inner_T_D} and \eqref{eq:inner_Omega_D} can be easily adjusted with weight matrices for non-uniform sampling. In the case of scattered data in space or time, the approximations \eqref{eq:inner_T_D} and \eqref{eq:inner_Omega_D} do not hold.

\subsection{RBF-based inner products}
\label{sec:2_1}

With the RBF regression, we seek to obtain an analytical expression of the velocity field $\boldsymbol{u}$ to approximate the inner products \eqref{inner_T} and \eqref{inner_Omega}. We assume that training data is available at randomly scattered locations in space and uniform points in time $\{\mathbf{t}_k\}^{n_t}_{k=1}$. While this structure is not a strict requirement of the method, it reflects the typical format of data obtained from particle tracking velocimetry and particle-based numerical simulations. To construct the analytical expression, we represent the velocity field at time $\mathbf{t}_i$ as a linear combination of $n_b$ RBFs:
\begin{equation}
    \label{eq:rbf_analytical_Velocity}
    \boldsymbol{u}(\boldsymbol{x}, \mathbf{t}_i) = \begin{pmatrix}
        u(\boldsymbol{x}, \mathbf{t}_i) \\[0.1cm]
        v(\boldsymbol{x}, \mathbf{t}_i)
    \end{pmatrix} \approx \begin{pmatrix}
        \boldsymbol{\gamma}^\T \big(\boldsymbol{x} ;\,\mathbf{X}_{c}(\mathbf{t}_i), \mathbf{c}(\mathbf{t}_i )\big) \mathbf{w}_{u} (\mathbf{t}_i) \\[0.1cm]
        \boldsymbol{\gamma}^\T \big(\boldsymbol{x} ;\,\mathbf{X}_{c}(\mathbf{t}_i), \mathbf{c}(\mathbf{t}_i )\big) \mathbf{w}_{v} (\mathbf{t}_i)
    \end{pmatrix}\,,
\end{equation} where $\boldsymbol{\gamma}\in\mathbb{R}^{n_b}$ is a vector stacking the value of $n_b$ radial basis functions $\{\gamma_m(\boldsymbol{x};\,\mathbf{X}_{c,m}(\mathbf{t}_i),\mathbf{c}_m(\mathbf{t}_i))\}_{m=1}^{n_b}$, at an arbitrary location $\boldsymbol{x}$. The set of basis function is identified by the vector of collocation points $\mathbf{X}_{c,m}(\mathbf{t}_i)$ and the vector of shape parameters $\mathbf{c}_m(\mathbf{t}_i)$, while the vectors $\mathbf{w}_u,\,\mathbf{w}_v \in \mathbb{R}^{n_b}$ gathers the weights of each RBF for each velocity component. In this work, we use radial basis function of the form:
\begin{align}
     \gamma_m (\boldsymbol{x} ;\, \mathbf{X}_{c,m}, \mathbf{c}_m) = \left(1 + \frac{\vert\vert \boldsymbol{x} - \mathbf{X}_{c,m} \vert\vert_2}{\mathbf{c}_m}\right)^5 \left(1 - \frac{\vert\vert \boldsymbol{x} - \mathbf{X}_{c,m} \vert\vert_2}{\mathbf{c}_m}\right)^5_+\,,
\end{align} where $\vert\vert \bullet \vert\vert_2$ is the $\ell_2$ norm of a vector, and the subscript $_+$ denotes the truncated positive part of a function, i.e. $(a)_+ = a$ if $a>0$ and $(a)_+=0$ if $a < 0$. For each snapshot, the weights can be computed via ridge regression \cite{Hastie2009, Bishop2011}, or through physics-constrained regression \cite{Sperotto2022a, Ratz2024}. More generally, enhancements such as spectral filtering, boundary condition penalties, or divergence free for coupled velocity-pressure fields can also be embedded in this step \cite{Gesemann2016, Jeon2022}, with linear constraints that could be implemented using Lagrange multipliers \cite{Sperotto2022a, Li2024, Ratz2024}. For the test cases investigated in this work, which consist of densely sampled two-dimensional slices of inherently three-dimensional datasets, the improvement in accuracy obtained by enforcing constraints was not sufficient to justify the associated increase in computational cost. We therefore adopt an unconstrained formulation here, while noting that such constraints can become highly beneficial in settings with lower spatial sampling density \cite{Sperotto2022a}. We use a compactly supported basis and solve the resulting least-squares problem with the iterative, sparse \texttt{lsmr} solver \cite{Fong2011} available in the python library \texttt{scipy}.

The analytic approximation in \eqref{eq:rbf_analytical_Velocity} can be used together with Gauss-Legendre quadrature or Monte Carlo methods to compute the inner products in \eqref{inner_Omega}, as proposed in \cite{Tirelli2025a}. In this work, we follow the alternative approach of computing the integral directly from the RBF approximation, taking the simplifying condition of having time independent collocation points and shape functions, i.e. $\boldsymbol{\gamma} (\boldsymbol{x} ;\, \mathbf{X}_{c}(\mathbf{t}_i), \mathbf{c}(\mathbf{t}_i) = \boldsymbol{\gamma} (\boldsymbol{x} ;\, \mathbf{X}_{c}, \mathbf{c})$. This assumption is valid if the domain boundaries are constant in time and if the data density is sufficiently uniform across the domain. Then, the inner product \eqref{inner_Omega} between two velocity snapshots at $\mathbf{t}_i$ and $\mathbf{t}_j$ becomes 

\begin{align}
    \label{eq:inner_product_rbf}
    \begin{split}&\hspace{-0.1cm} \left\langle \boldsymbol{u}(\boldsymbol{x}, \mathbf{t}_i),\, \boldsymbol{u}(\boldsymbol{x}, \mathbf{t}_j) \right\rangle_\Omega \approx \langle \boldsymbol{u}(\boldsymbol{x}, \mathbf{t}_i),\, \boldsymbol{u}(\boldsymbol{x}, \mathbf{t}_j)\rangle_{\Omega,a}  \\
    &\hspace{0.05cm} = \frac{1}{\vert \Omega \vert} \bigintssss_{\Omega} \sum_{k=u,v} \left( \sum_{m=1}^{n_b} \boldsymbol{\gamma}_m (\boldsymbol{x}) \mathbf{w}_{l,m}(\mathbf{t}_i) \right) \left( \sum_{n=1}^{n_b} \boldsymbol{\gamma}_n^\ast (\boldsymbol{x}) \mathbf{w}_{l,n}^\ast(\mathbf{t}_j) \right) \text{d}\Omega \\
    &\hspace{0.05cm} = \sum_{k=u,v} \sum_{m=1}^{n_b} \sum_{m=1}^{n_b} \mathbf{w}_{l,m}(\mathbf{t}_i) \mathbf{w}_{l,n}^\ast(\mathbf{t}_j) \underbrace{ \left( \frac{1}{\vert \Omega \vert} \bigintssss_{\Omega} \boldsymbol{\gamma}_m^\ast(\boldsymbol{x}) \boldsymbol{\gamma}_n(\boldsymbol{x}) \text{d}\Omega \right)}_{ \coloneqq \mathbf{I}_\Omega} \\
    &\hspace{0.05cm} = \sum_{k=u,v} \mathbf{w}_k^\ast(\mathbf{t}_i) \, \mathbf{I}_\Omega \, \mathbf{w}_k^{\vphantom{\ast}}(\mathbf{t}_j)\,,
    \end{split}
\end{align} having defined the RBF based inner product $\langle \bullet, \bullet \rangle_{\Omega,a}$ and having introduced the integrating matrix $\mathbf{I}_\Omega\in \mathbb{R}^{n_b \times n_b}$ gathering the inner products of the basis functions.

The extension to inner products in time is analogous, relaxing the standard assumption of uniform temporal sampling and enabling decompositions that are continuous in both space and time. This work focuses on the discrete-time formulation, but we highlight that spatial RBF regression yields continuous spatial structures. The following sections describe the meshless implementation of the key decomposition methods.

\subsection{Meshless algorithms}
\label{sec:2_3}

We introduce the meshless variant of the traditional POD (Sec. \ref{sec:2_3_1}) and its spectrally constrained variants (Sec. \ref{sec:2_3_2}), the Spectral POD (Sec. \ref{sec:2_3_3}) and the Dynamic Mode Decomposition (Sec. \ref{sec:2_3_4}). The extension of other grid-based decomposition methods to the RBF-induced meshless framework is conceptually straightforward and can be carried out following analogous steps.

\subsubsection{Proper orthogonal decomposition (POD) }
\label{sec:2_3_1}

In the classic POD, the temporal structures of the modes are eigenfunction of the two point correlation function. The formulation for the continuous problem is usually presented for the so called ``space-only" POD \cite{Berkooz1993,Towne2018,Holmes2012}, but a similar derivation applies for the time domain. 

The temporal structures $\psi_r(t)$ for the POD are taken to optimally represent the evolution of the field $\boldsymbol{u}(\boldsymbol{x},t)$ at any location $\boldsymbol{x}$. To obtain these modes, we define the temporal correlation kernel as 
\begin{equation}
K(t,t') = \int_{\Omega} \boldsymbol{u}(x,t) \boldsymbol{u}(x,t') \, \text{d}\boldsymbol{x}\,,
\end{equation} and the associated self-adjoint operator  $ \mathcal{K} $  acting on functions $ \psi \in L^2([0,T]) $ via:
\begin{equation}
(\mathcal{K} \psi)(t) = \int_0^T K(t,t')\,\psi(t') \, \text{d}t'\,.
\end{equation}

With these definitions, the optimal temporal basis functions $\psi(t)$, which minimize the $l_2$ norm of the approximation error for any number of selected modes, are the solutions to the Fredholm integral eigenvalue problem:

\begin{equation}
\label{eig_C}
\int_0^T K(t,t')\,\psi_r(t') \, \text{d}t' = \lambda_r \psi_r(t)\,.
\end{equation}

These bases are orthogonal under the continuous inner product \eqref{inner_T}. In the traditional setting of discrete time realizations available on a uniform grid $\{\mathbf{t}_k\}^{n_t}_{k=1}$, the continuous eigenvalue problem \eqref{eig_C} is turned into a matrix eigenvalue problem for the time correlation matrix $\mathbf{K}_{i,j}=\langle \boldsymbol{u}(\boldsymbol{x}, \mathbf{t}_i),\,\boldsymbol{u}(\boldsymbol{x}, \mathbf{t}_j) \rangle_{\Omega,d}$. This matrix collects the inner product in space between all the available pairs of snapshots.  Since $\mathbf{K}\in\mathbb{R}^{n_t\times n_t}$ is at least positive semidefinite by definition, the eigenvalue problem reads 
\begin{align}
\label{eq:eig_K}
	\mathbf{K} = \boldsymbol{\Psi} \boldsymbol{\Lambda} \boldsymbol{\Psi}^\T\,,
\end{align} and the eigenvectors $\{\boldsymbol{\Psi}_r\}^{n_t}_{r=1}$ of $\boldsymbol{\Psi}\in\mathbb{R}^{n_t\times n_t}$ can be interpreted as discrete approximations, evaluated on the uniform time grid $\mathbf{t}$, of the continuous temporal eigenfunctions $\psi_r(t)$. These vectors are orthogonal under the inner product \eqref{eq:inner_T_D}.

The proposed meshless variant of the POD replaces the inner product in the definition of $\mathbf{K}$ with the RBF-based inner product in \eqref{eq:inner_product_rbf}, i.e. $\mathbf{K}_{ij} \approx \langle \boldsymbol{u}(\boldsymbol{x}, \mathbf{t}_i),\,\boldsymbol{u}(\boldsymbol{x}, \mathbf{t}_j) \rangle_{\Omega,a}$, thus allowing to handle scattered data. In the simpler setting of fixed RBF bases the integrating matrix $\mathbf{I}_\Omega$ is independent of time, and the temporal correlation matrix can be computed as:
\begin{align}
    \label{eq:K_rbf_mat}
    \mathbf{K} = \sum_{k=u,v} \mathbf{W}_k^\T \mathbf{I}_\Omega\,\mathbf{W}_k^{\vphantom{T}}\,,
\end{align} where $\mathbf{W}_u,\, \mathbf{W}_v \in \mathbb{R}^{n_b \times n_t}$ collect the weights of all snapshots for each velocity component. Note that, with a slight abuse of notation, we use the same symbols $\mathbf{K}$ and $\boldsymbol{\Psi}$ to refer to the temporal correlation matrix and its eigenvectors, regardless of whether $\mathbf{K}$ is computed on grid data using the inner product \eqref{eq:inner_Omega_D} between snapshots, or via the RBF-based inner product between regressions. The distinction between the two is made clear in the following when strictly necessary.

We stress that the definition \eqref{eq:K_rbf_mat} also leads to a matrix that is at least positive semidefinite, hence having orthogonal eigenvectors under the inner product \eqref{eq:inner_T_D}.

The second step in the traditional POD computation, following Sirovich's method \citep{Sirovich1987}, is the projection of the data onto the bases $[\boldsymbol{\Psi_r}]^{n_R}_{r=1}$ either using the continuous inner product $\langle \rangle_{T}$ in \eqref{inner_T} for continuous data or the discrete one $\langle \rangle_{T,d}$ in \eqref{eq:inner_T_D} for gridded data. Since the orthogonality of the temporal structure $\boldsymbol{\Psi}_r$ is preserved by construction, the same approach remains valid for the meshless formulation proposed in this work. Therefore, it is possible to obtain an approximation of the continuous spatial structure for the $u$ and $v$ components associated to the $r$-th temporal structure, denoted as $(\phi_{u,r}(\boldsymbol{x})$ and $\,\phi_{v,r}(\boldsymbol{x})$ respectively, with the discrete inner product in time. For the velocity component $u$, for example, this reads:
\begin{align}
    \begin{split}
    \label{eq:rbf_spatial_structures}
   \sigma_r \mathrm{\phi}_{u,r}(\boldsymbol{x}) &= \left\langle {u}(\boldsymbol{x}, t),\,\boldsymbol{\psi}_r(t) \right\rangle_T\\
    &\approx \left\langle {u}(\boldsymbol{x}, \mathbf{t}_i),\,\boldsymbol{\Psi}_r(\mathbf{t}_i) \right\rangle_{T,d}\\
    &\approx \frac{1}{ n_t} \boldsymbol{\gamma}^\T(\boldsymbol{x}) \sum_{i=1}^{n_t}\mathbf{w}_u(\mathbf{t}_i) \, \boldsymbol{\Psi}_r(\mathbf{t}_i)\\
    &= \frac{1}{n_t} \boldsymbol{\gamma}^\T(\boldsymbol{x}) \bigl(\mathbf{W}_u \boldsymbol{\Psi}_r\bigr) = \frac{1}{n_t} \boldsymbol{\gamma}^\T(\boldsymbol{x}) \mathbf{w}_{\phi,u,r}   \,,
    \end{split} 
\end{align} where $\sigma_r$ is the $r$-th modal amplitude, $\boldsymbol{\psi}_r(\mathbf{t}_i)$ is the $i$-th sample of the $r$-th temporal structure, $\boldsymbol{\gamma}^\T(\boldsymbol{x})$ is the shortened notation for $\boldsymbol{\gamma}^\T (\boldsymbol{x} ;\, \mathbf{X}_{c}, \mathbf{c})$ and $\mathbf{w}_{\phi,u,r}$ is the weight vector for the RBF approximation of the u component of the $r$-th spatial structure $\boldsymbol{\phi}_r(\boldsymbol{x})=[\phi_{u,r}(\boldsymbol{x}),\phi_{v,r}(\boldsymbol{x})]$.

We note that the amplitude $\sigma_r$ serves as a normalization factor to ensure unitary norm of the spatial structures under a given inner product. A natural choice is to set

\begin{align}
    \sigma_r^2 &= \left\langle \, \langle  \boldsymbol{u}(\boldsymbol{x}, t),\, \mathbf{\Psi}_r(t) \rangle^{\vphantom{T}}_{T,d} \, \right\rangle^2_\Omega\,.
\end{align} 

In the case of gridded data, where POD effectively reduces to an SVD of the snapshot matrix \cite{Mendez2023}, one has $\sigma_r = \lambda_r^2$, with $\lambda_r$ the eigenvalues in \eqref{eq:eig_K}. This relationship arises from the symmetry of the inner product applied to both the column and row spaces of the snapshot matrix. In the proposed RBF-based formulation, however, this symmetry no longer holds: projections in time are defined under the inner product \eqref{eq:inner_T_D} while projection in space are defined under the inner product \eqref{eq:inner_product_rbf}. Moreover, the temporal modes are orthogonal under \eqref{eq:inner_T_D} while the spatial ones are not orthogonal under \eqref{eq:inner_product_rbf}, since:

\begin{equation}
\begin{split}
\label{phi_check}
\langle \boldsymbol{\phi}_r\,,\, \boldsymbol{\phi}_s\rangle_{\Omega,a}&\propto\int_{\Omega} \Biggl(\sum^{n_b}_{l=1} w_{r,l} \gamma_l(\boldsymbol{x}) \Biggr) \Biggl(\sum^{n_b}_{l=1} w_{s,l} \gamma_l(\boldsymbol{x}) \Biggr) d \Omega\\&=\mathbf{w}_{\phi,r}^T \mathbf{I}_{\Omega} \mathbf{w}_{\phi,s}\,,
\end{split}
\end{equation} where $\mathbf{w}_{\phi,i}$ and $\mathbf{w}_{\phi,j}$ are the RBF weights for the spatial structures following the time projection in \eqref{eq:rbf_spatial_structures}. Thus we see that preserving the spatial orthogonality of the POD modes under \eqref{eq:inner_product_rbf} requires that the weight vectors of the RBF expansions of the spatial modes be orthogonal with respect to the inner product weighted by the integration matrix $\mathbf{I}_{\Omega}$.  This condition could be enforced as an additional constraint during the RBF regression of all snapshots, using an iterative procedure that alternates between computing the temporal correlation matrix $\mathbf{K}$, its eigenvectors $\boldsymbol{\Psi}$, the associated RBF weights $\mathbf{w}_{\phi,i}$, and the inner products in \eqref{phi_check}. Nevertheless, since the loss of orthogonality was found to be minor, we leave iterative correction as a topic for future work. We note, however, that such an extension would become particularly relevant when constructing meshless methods for Galerkin-based reduced-order models.

\subsubsection{Spectrally constrained PODs: the SPOD and mPOD}\label{sec:2_3_2}
The Spectral POD by \citet{Sieber2016} (hereafter referred to as \textit{Sieber SPOD}) and the Multiscale POD by \citet{Mendez2019} (hereafter \textit{mPOD}) follow the same general steps as the traditional snapshot POD, with additional operations applied to the temporal correlation matrix $\mathbf{K}$. Specifically, Sieber SPOD applies a filter along the diagonals of $\mathbf{K}$ to promote a circulant structure, yielding eigenvectors that more closely resemble harmonic modes. The mPOD employs multiresolution analysis to decompose $\mathbf{K}$ into a sum of components at different temporal scales, each of which is equipped with its own POD. The reader is referred to the original articles for further details. The extension of these algorithms to the RBF-based meshless formalism introduces no additional steps beyond those already required for the traditional POD.

\subsubsection{A harmonic POD: another SPOD}
\label{sec:2_3_3}

The SPOD of \citet{Towne2018} (hereafter referred to as `Towne SPOD') uses Welch's method \cite{Welch1967}, and partitions the data in time into $n_B$ different overlapping blocks of length $n_f$. A discrete Fourier transform (DFT) is carried out on each of these blocks. For the proposed meshless RBF based variance, the DFT of the data simply turns into a DFT of the weights if the RBF basis is fixed in time: 
\begin{align}
    \hat{\boldsymbol{u}}(\boldsymbol{x}, \mathbf{f}_l) = \begin{pmatrix}
        \hat{u}(\boldsymbol{x}, \mathbf{f}_l) \\
        \hat{v}(\boldsymbol{x}, \mathbf{f}_l)
    \end{pmatrix} \approx \begin{pmatrix}
        \boldsymbol{\gamma}^\T(\boldsymbol{x}) \mathbf{W}_u \, \boldsymbol{\Psi}_\mathcal{F} \\
        \boldsymbol{\gamma}^\T(\boldsymbol{x}) \mathbf{W}_v \, \boldsymbol{\Psi}_\mathcal{F}
    \end{pmatrix}
    = \begin{pmatrix}
        \boldsymbol{\gamma}^\T(\boldsymbol{x}) \hat{\mathbf{W}}_u \\
        \boldsymbol{\gamma}^\T(\boldsymbol{x}) \hat{\mathbf{W}}_v
    \end{pmatrix} \,,
\end{align} where $\boldsymbol{\Psi}_\mathcal{F}\in\mathbb{C}^{n_f\times n_f}$ is the Discrete Fourier Transform matrix \cite{Golub2013}, defined as $\boldsymbol{\Psi}_{\mathcal{F},l}(\mathbf{t}_k)=\mbox{exp}(2\pi \mathbf{f}_l \mathbf{t}_k \mathrm{i})/\sqrt{n_f}$ with $\{\mathbf{f}_l= l/n_f f_s\}^{n_f}_{l=1}$ the set of uniformly spaced frequencies and $\hat{\mathbf{W}}_u$ and $\hat{\mathbf{W}}_v$ the row-wise Fourier transforms of the weight matrices. For every (discrete) frequency in the set $\{\mathbf{f}_l\}^{n_f}_{l=1}$, the DFTs of each block are then stacked together. For the velocity $u$, for example, this results in:
\begin{align}
    \begin{split}
        \hat{\mathbf{u}}_\mathcal{W}(\boldsymbol{x}, \mathbf{f}_l) &= \left[ \hat{{{u}}}^{(1)}(\boldsymbol{x},\mathbf{f}_l),\, \dots,\, \hat{{{u}}}^{(n_B)}(\boldsymbol{x},\mathbf{f}_l) \right] \\
        &\approx  \boldsymbol{\gamma}^\T(\boldsymbol{x}) \left[ \hat{\mathbf{{w}}}_u^{(1)}(\mathbf{f}_l),\, \dots,\, \hat{\mathbf{{w}}}_u^{(n_B)}(\mathbf{f}_l) \right]\\
        &=  \boldsymbol{\gamma}^\T(\boldsymbol{x}) \hat{\mathbf{{W}}}_{\mathcal{W},u}(\mathbf{f}_l)\,,
    \end{split}
\end{align} where $\hat{\boldsymbol{{w}}}_u^{(p)}(\mathbf{f}_l)$ collects the DFT of the $p$-th Welch block at the $l$-th frequency and $\hat{\mathbf{{W}}}_{\mathcal{W}, u} \in \mathbb{C}^{n_b \times n_B}$ gathers the individual Welch contributions at frequency $\mathbf{f}_l$ of the Fourier transform of the weights, as in the original work \cite{Towne2018}.

This Welch block is formed for both velocity components and used to form a correlation matrix $\mathbf{M} \in \mathbb{C}^{n_B \times n_B}$ with an inner product in space:
\begin{align}
    \begin{split}
        \mathbf{M} (\mathbf{f}_l) &= \left\langle \hat{\boldsymbol{u}}(\boldsymbol{x}, \mathbf{f}_l),\, \hat{\boldsymbol{u}}(\boldsymbol{x}, \mathbf{f}_l)^{\vphantom{\ast}} \right\rangle_{\Omega,a} \\
        &= \sum_{k=u,v} \hat{\mathbf{W}}^\ast_{\mathcal{W}, k}(\mathbf{f}_l) \, \mathbf{I}_\Omega\, \hat{\mathbf{W}}_{\mathcal{W}, k}(\mathbf{f}_l)^{\vphantom{\ast}}\,,
    \end{split}
\end{align} where the derivation from the inner product to the integration matrix $\mathbf{I}_\Omega$ follows the same steps as \eqref{eq:inner_product_rbf}.

As in the original work, an eigenvalue decomposition is then carried out on this matrix for each frequency, that is $\mathbf{M}(\mathbf{f}_l) = \boldsymbol{\Theta}(\mathbf{f}_l) \, \boldsymbol{\Sigma}^2(\mathbf{f}_l) \, \boldsymbol{\Theta}^\ast(\mathbf{f}_l)$, and the $r$-th spatial mode of $u$ associated to frequency $\mathbf{f}_l$ is obtained via
\begin{align}
    {\phi}_{u,r}(\boldsymbol{x}, \mathbf{f}_l) = \frac{1}{\sigma_r(\mathbf{f}_l)} \boldsymbol{\gamma}^\T (\boldsymbol{x}) \, \hat{\mathbf{{W}}}_{\mathcal{W},u}(\mathbf{f}_l) \, \boldsymbol{\Theta}_r(\mathbf{f}_l)\,,
\end{align} where $\boldsymbol{\Theta}_r(\mathbf{f}_l)$ is the $r$-th column of $\boldsymbol{\Theta}(\mathbf{f}_l)$. The result is a total of $n_B$ modal amplitudes and spatial structures at each frequency $\mathbf{f}_l$.


\subsubsection{Dynamic Mode Decomposition (DMD)}
\label{sec:2_3_4}

Many algorithms for computing the DMD have been proposed \cite{Tu2014}. We here consider the POD-based formulation proposed by \citet{Schmid2010} which seeks to identify the eigenvalue decomposition of a reduced propagator $\tilde{\boldsymbol{S}}$ that best approximates, in a least square sense, the dynamics in the POD-reduced space using a linear system. In the traditional grid based formulation, denoting as $\boldsymbol{D}_1$ and $\boldsymbol{D}_2\in\mathbb{R}^{n_s\times (n_t-1)}$ the snapshot matrices containing the snapshots from $1$ to $n_t-1$ and from $2$ to $n_t$ and as $\boldsymbol{P}\in\mathbb{R}^{n_s\times n_s}$ the (unknown) full propagator such that $\boldsymbol{D}_2=\boldsymbol{P}\boldsymbol{D}_1$, the reduced propagator is defined as $\tilde{\boldsymbol{S}}=\boldsymbol{\tilde{\Phi}}_\mathcal{P}^{\T}\boldsymbol{P\boldsymbol{\tilde{\Phi}}}_\mathcal{P}$, where $\sim$ denotes quantities computed from a reduced number of modes. Using the truncated POD (SVD) to compute the pseudoinverse $\boldsymbol{D}^{\dagger +}=\tilde{\boldsymbol{\Psi}}_\mathcal{P} \tilde{\boldsymbol{\Sigma}}_\mathcal{P} \tilde{\boldsymbol{\Phi}}^{-1}_\mathcal{P}$, and introducing it in the least square definition of the propagator $\boldsymbol{P}= \boldsymbol{D}_2 \tilde{\boldsymbol{\Psi}}_\mathcal{P} \tilde{\boldsymbol{\Sigma}}_\mathcal{P} \tilde{\boldsymbol{\Phi}}^{-1}_\mathcal{P}$, the reduced propagator becomes \cite{Schmid2010, Tu2014}:

\begin{equation}
\label{eq:reduced_propagator}
\tilde{\boldsymbol{S}}=\boldsymbol{\tilde{\Phi}}^{\T}_{\mathcal{P}} \boldsymbol{D}_2 \boldsymbol{\tilde{\Psi}}_{\mathcal{P}} \boldsymbol{\tilde{\Sigma}}^{-1}_{\mathcal{P}},
\end{equation}

The meshless variant of \eqref{eq:reduced_propagator} can be easily derived by replacing $\tilde{\boldsymbol{\Sigma}}_\mathcal{P}$ and $\tilde{\boldsymbol{\Psi}}_\mathcal{P}$ with their meshless counterpart and replacing the inner product $\tilde{\boldsymbol{\Phi}}^T_\mathcal{P} \boldsymbol{D}_1$ with the RBF-based one.

A simpler and computationally cheaper alternative is offered by variants of this idea that were developed earlier in atmospheric science, most notably the Principal Oscillation Pattern (POP) and Principal Interaction Pattern (PIP) frameworks \cite{Storch1990,Penland1996,Hasselmann1988}, which also estimate a linear propagator directly from time-lagged snapshots.

 These allow to compute the reduced propagator without involving inner product in the space domain. The underlying idea is to assume that the two shifted snapshot sequences (the snapshot matrices $\boldsymbol{D}_1,\boldsymbol{D}_2$ in traditional DMD, or their space continuous formulation in the meshless approach) share the same POD and only differ in the temporal structures. Then, it can be shown (see \cite{Mendez_2025_book2}) that  \eqref{eq:reduced_propagator} becomes: 

\begin{equation}
\label{reduced_Prop_2}
    \tilde{\boldsymbol{S}} = \tilde{\boldsymbol{\Sigma}}_\mathcal{P}^{-1} \tilde{\boldsymbol{\Psi}}_{\mathcal{P},1}^\T \tilde{\boldsymbol{\Psi}}_{\mathcal{P},2} \tilde{\boldsymbol{\Sigma}}_\mathcal{P},
\end{equation} where $\boldsymbol{\Psi}_{\mathcal{P},1} = \{\boldsymbol{\Psi}_\mathcal{P}(\mathbf{t}_i)\}_{i=1}^{n_t-1}$ and $\boldsymbol{\Psi}_{\mathcal{P},2} = \{\boldsymbol{\Psi}_\mathcal{P}(\mathbf{t}_i)\}_{i=2}^{n_t}$, with $\boldsymbol{\Psi}_\mathcal{P}$ the temporal structure of the data POD and $\tilde{\boldsymbol{\Sigma}}_\mathcal{P}$ the associated amplitude matrix. The meshless variant of \eqref{reduced_Prop_2} follows directly from the meshless POD construction and requires no additional RBF inner products beyond the single POD computation.

Finally, as for the grid-based DMD, the eigenvalue decomposition of the reduced propagator is carried out $\tilde{\boldsymbol{S}} = \mathbf{V} \mathbf{\Lambda} \mathbf{V}^\ast$ and the resulting spatial structures are computed from spatial projection of the POD structure. For the meshless DMD, this projection is computed from the RBF weights, the POD of the data, and the eigenvectors of the reduced propagator. For the u-component of the spatial structure, for example, one has:
\begin{align}
    \begin{split}
        {\phi}_{u,r}(\boldsymbol{x}) &=  \tilde{\boldsymbol{\phi}}_\mathcal{P}(\boldsymbol{x}) \,\mathbf{q}_r = \frac{1}{\sigma_r n_t} \boldsymbol{\gamma}^\T(\boldsymbol{x}) \mathbf{W}_u \mathbf{\Psi}_{\mathcal{P},r} \mathbf{V}_r\,.
    \end{split}
\end{align}

\section{Selected test cases}
\label{sec:3}

\begin{figure}[t]
    \centering
    \includegraphics[width=\linewidth]{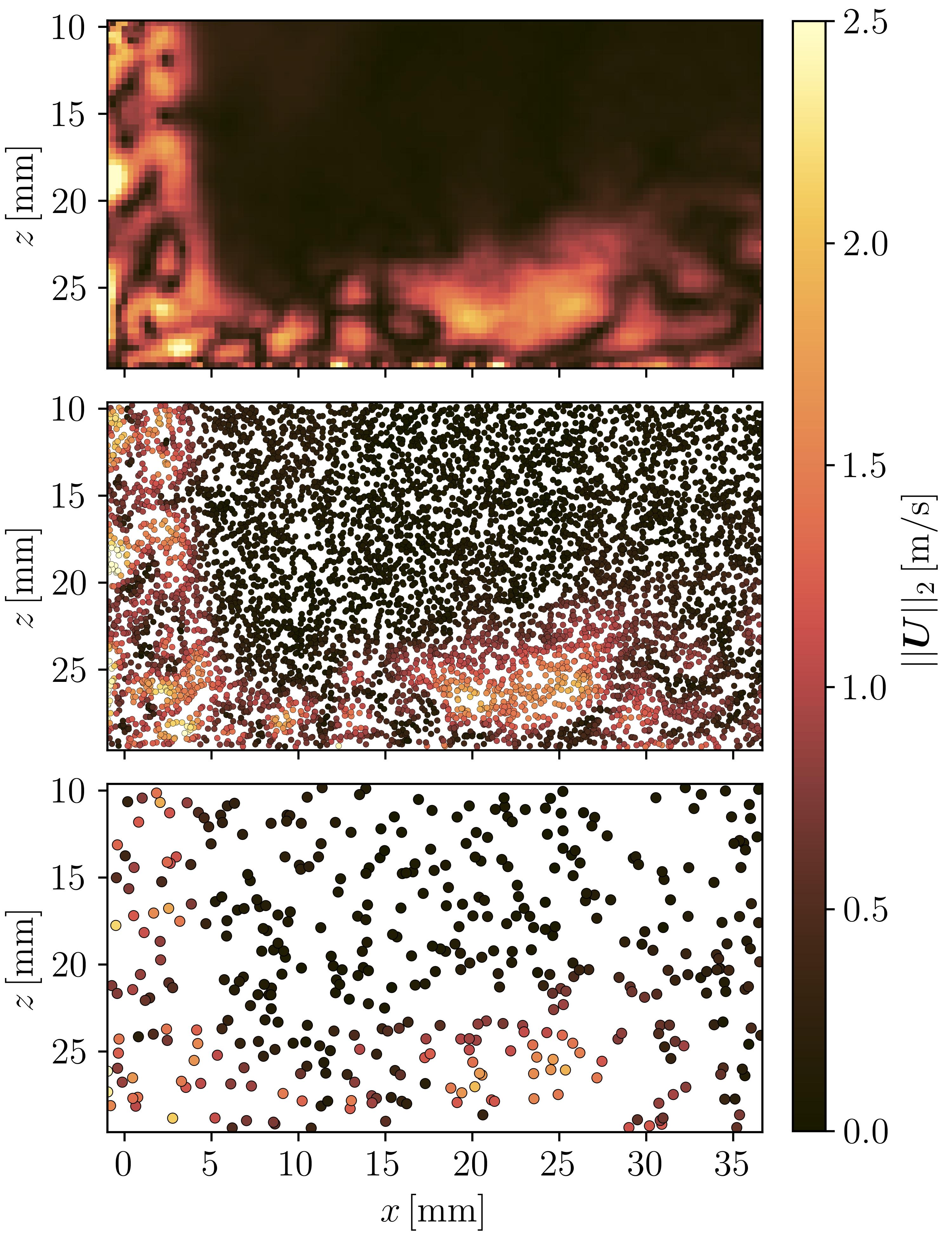}
    \caption{Test case 1. Mean subtracted velocity magnitude for the impinging jet. Example PIV snapshot (top) and the time step interpolated onto scattered points with downsampling factor $k=1$ (middle) and $k=16$ (bottom).}
    \label{fig:3_1_testcase_jet}
\end{figure}

The meshless decompositions are analyzed using two different datasets. The first dataset contains experimental data from PIV and features two typical experimental challenges: The spatial resolution is coarse because of the low-pass filtering of the interrogation windows, and the range of relevant frequencies in time is close to the Nyquist limit. This is common for experiments where increasing the spatial or temporal resolution arbitrarily is limited by hardware. The second dataset contains numerical data from large eddy simulations (LES) and features the contrary: The data are temporally well resolved to capture relevant temporal frequencies and contains spatial scales down to the filtering frequency of the LES model.

For both datasets, we obtain the scattered datasets by interpolating the original (gridded) data. We interpolate the data with different downsampling factors $k$, where $n_p = n_g / k$, simulating different data densities. Before interpolating, we remove the mean of the gridded data, since both datasets are stationary. If the gridded mean is unavailable, it can be subtracted from the data ensemble through binning \citep{Tirelli2023b}, or another RBF regression \citep{Ratz2024}.

\subsection{Test case 1: Impinging jet from particle image velocimetry}
\label{sec:3_1}

The first test case uses experimental data from PIV of a jet impinging on a plate, denoted as `Jet' in the remainder of the manuscript. The jet flow features three distinct flow patterns: the nozzle exit, the impinging region close to the wall, and the developing wall jet. Each of these occurs at a different spatial location, with a different frequency in time and in space. The measurements were conducted at the von Karman Institute and featured a rectangular nozzle at a distance of $Z = \SI{30}{\milli\meter}$ with a size of $H \times W = 4 \times \SI{250}{\milli\meter}$ and an exit velocity of $U_0 = \SI{6.5}{\meter\per\second}$, corresponding to a Reynolds number of \num{1733}. Further details can be found in \citet{Mendez2019}. The domain of interest had a size of $20 \times \SI{38}{\milli\meter}^2$, and \num{2000} snapshots were acquired at a frequency of \SI{2}{\kilo\hertz}. This corresponds to a dimensionless time $t\,U_0/ H$ in the range $[0, 1625]$ with a temporal resolution $\Delta t\,U_0/ H = 0.81$. Relevant coherent structures are present up to a dimensionless frequency of approximately $\Strouhal = 0.4$, with $\Strouhal$ the Strouhal number, while the Nyquist frequency is $\Strouhal=f H / U_0=0.61$. 

The resulting velocity is available in $n_g = 6840$ points on a regular grid of $60 \times 114$ in $z$ and $x$-direction with a vector pitch of $\Delta x = \SI{0.333}{\milli\meter}$. Since the grid is regular, we use piecewise cubic Hermite interpolating polynomials to produce scattered data. Three different datasets are generated with different downsampling factors $k \in [1, 4, 16]$. The first panel of Figure~\ref{fig:3_1_testcase_jet} shows the (mean-subtracted) magnitude of a PIV snapshot, illustrating the different turbulent length scales in the free shear layer and the developing wall jet. The bottom two subpanels show the same snapshot interpolated with $k=1$ and $k=16$. For the latter, the different vortices are hardly visible by eye.

\subsection{Test case 2: Transitional airfoil from large eddy simulations}
\label{sec:3_2}

\begin{figure}[tb]
    \centering
    \includegraphics[width=\linewidth]{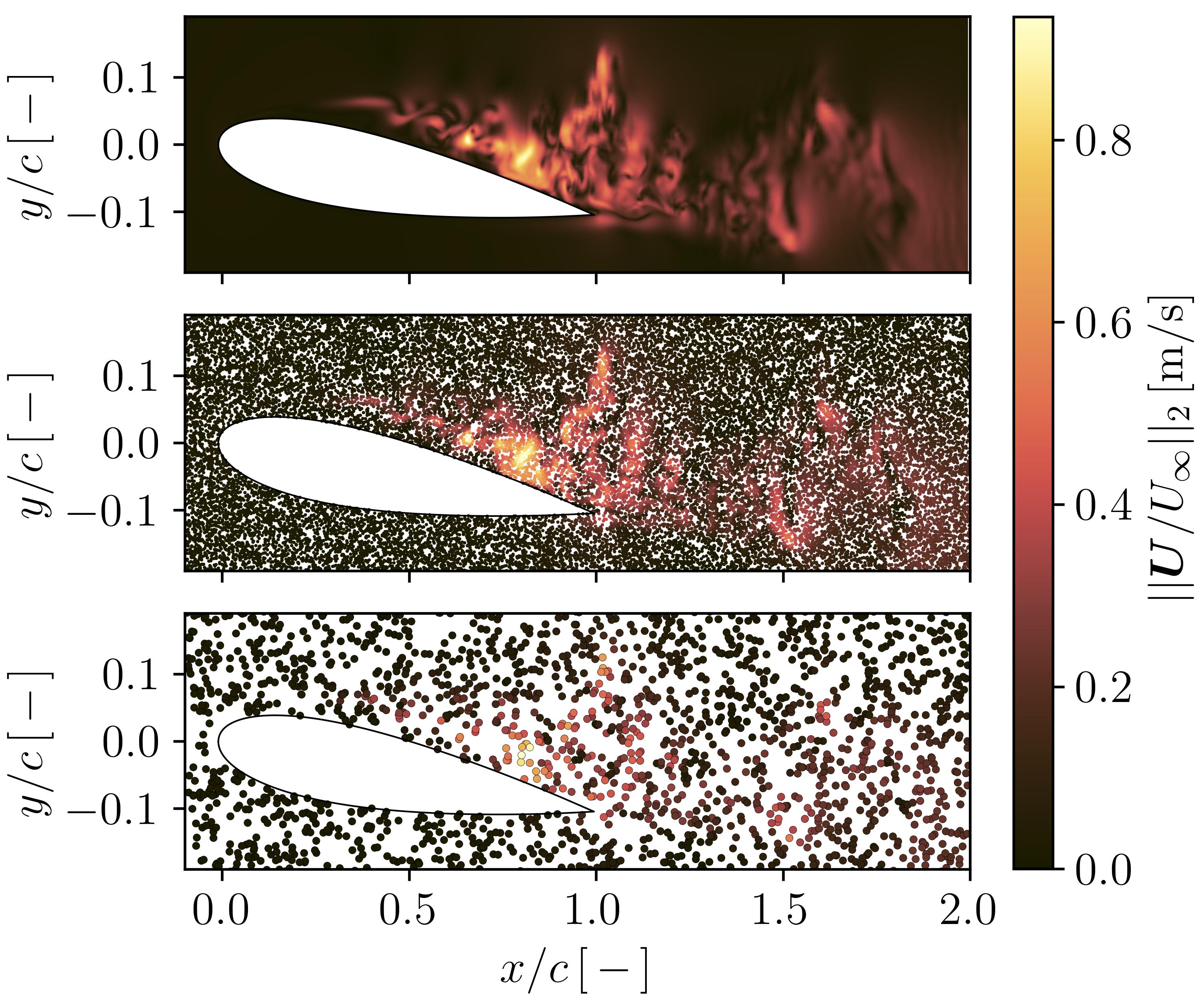}
    \caption{Test case 2. Mean subtracted velocity magnitude for the airfoil flow. Example LES snapshot (top) and the time step interpolated onto scattered points with downsampling factor $k=2$ (middle) and $k=32$ (bottom). The aspect ratio is slightly increased for better visualization.}
    \label{fig:3_2_testcase_airfoil}
\end{figure}

The second test case uses data from an LES of the flow past a NACA 0012 at \SI{6}{\degree} angle of attack at a Reynolds number of \num{23000}, denoted as `Airfoil' in the remainder. In these conditions, the laminar flow separates near the leading edge, transitions to turbulence, and reattaches close to the trailing edge. The Kelvin-Helmholtz instability triggers vortex shedding over the separation bubble, leading to a wide spectrum over the airfoil and in the wake. Further details can be found in \citet{Yeh2019, Towne2023}. From the full simulation environment, we extract data in a domain $(x/c,\, y/c) \in [-0.1, 2] \times [-0.2, 0.2]$, where $c$ is the chord of the airfoil. The data are available in \num{16000} snapshots spanning a dimensionless time of $t\, c / U_\infty \in [0, 165]$ corresponding to a dimensionless timestep of $\Delta t\, c / U_\infty = 0.0104$. Relevant structures appear up to a frequency of $\Strouhal \approx 4$, giving this test case a much higher temporal resolution with a Nyquist frequency of $\Strouhal = 48$.

We sample span-wise averaged velocities $u$ and $v$ in the $n_g = \num{60844}$ unstructured LES points. Since this grid is irregular, we resort to an RBF interpolation with a linear kernel to avoid ill-conditioning caused by the hole in the domain. To avoid excessive oversampling in regions with large grid spacing (i.e., far from the airfoil), we interpolate this data with the three downsampling factors $k \in [2, 8, 32]$. The first panel of Figure~\ref{fig:3_2_testcase_airfoil} shows the (mean subtracted) velocity magnitude, which displays the characteristics of the laminar separation bubble over the chord and the resulting turbulent wake. The lower panels of the same figure show the same time step, interpolated with $k=2$ and 32. For the dense case, the turbulent structures are again clearly visible over the airfoil and in the wake. For $k=32$, only \num{1900} points are available, which makes it almost impossible to spot structures in the wake or over the latter half of the chord.

\begin{figure}[tb]
    \centering
    \includegraphics[width=\linewidth]{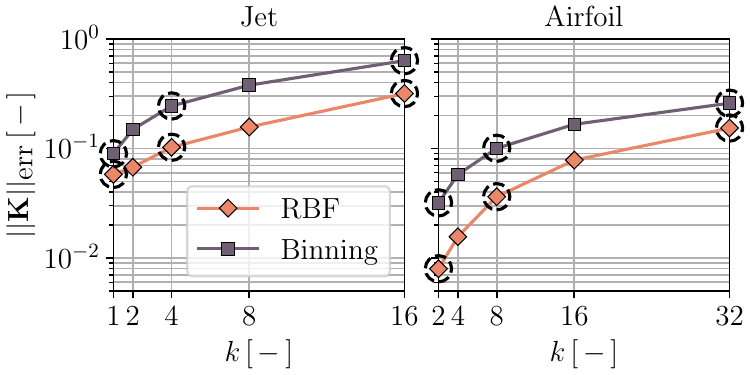}
    \caption{Normalized $L_2$ error in the temporal correlation matrix $\mathbf{K}$ \eqref{eq:k_err} versus the downsampling factor for binning and RBFs for the jet (left) and the airfoil (right). The dashed circles correspond to the downsampling factors analyzed in detail in the remainder of the article.}
    \label{fig:4_1_K_convergence}
\end{figure}


\begin{figure*}[t]
    \centering
    \includegraphics[width=\linewidth]{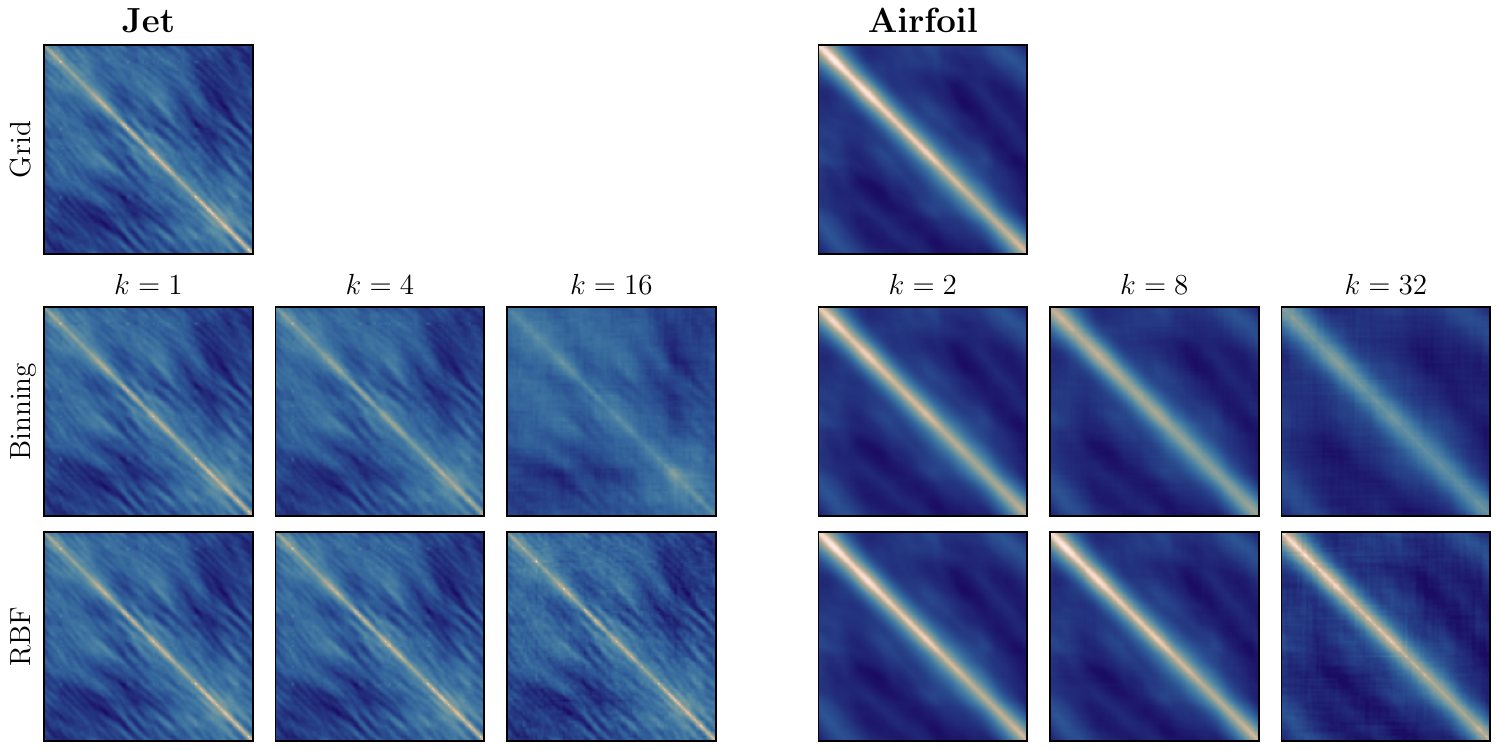}
    \caption{Temporal correlation matrix $\mathbf{K}$ of the jet data (left) and airfoil data (right). The upper left panel on each side shows the gridded (ground truth) data while the second and third row respectively show the binning and RBF results. The columns on each side show different downsampling factors, and the colorscale is shared across all subpanels. For better visualization, only the first 200 time steps of each dataset are displayed.}
    \label{fig:4_2_K_matrices}
\end{figure*}

\section{Results}
\label{sec:4}

We compare the RBF-based modal decompositions with the reference data and binning. For the jet flow, the gridded PIV is taken as the ground truth, acknowledging the data being smoothed and contaminated by noise. The spatial structures are visualized in the grid points of the underlying data, meaning the RBFs are evaluated in these points and binning uses them as centroids. The bin size is chosen such that each bin contains 10 particles on average; a trade-off between accuracy and noise filtering, used in other benchmarks \cite{Sciacchitano2025}. Finally, the sample-based inner products of binning and the ground truth of the airfoil are computed with a weighted inner product based on the cell volume \cite{Mendez2023}.

For both datasets, the RBF regression uses the same parameters. We choose a ratio between training points and RBFs of $n_p / n_b = 1.5$ and select a uniform shape factor $c$ such that an average of 40 data points is underneath each RBF. They are placed semi-randomly in space with Halton points, and collocation points are removed within the airfoil to mitigate ill-conditioning. The ridge regression penalty is set to correspond to $\alpha_\textrm{ridge} = 0.01$ for a direct least squares solver \cite{Hastie2009, Bishop2011}. These hyperparameters are selected based on similar regressions (e.g. \cite{Gesemann2016, Sperotto2024b}), but we observed the results to depend relatively little on the hyperparameters. A reasonable regression always resulted in reasonable decompositions.

For the Towne SPOD, we use Welch blocks with \SI{50}{\percent} overlap with $N_f = 200$ and 2000 samples for the jet and the airfoil. The results likewise varied little for any reasonable range of overlaps. For the Sieber SPOD and mPOD, we respectively tested different filter cut-offs and filter banks, and we show results sparingly where appropriate. Finally, we compute the DMD with 200 and 2000 POD modes for the jet and the airfoil.

The results are presented not split by the different methods, but by their common steps. That is, we start with the correlation matrix based on the spatial inner product and then continue with the decomposition of this matrix into modal amplitudes and the temporal bases. From there, we obtain the spatial structures, either by projecting the data onto the temporal basis structures (POD/mPOD/SPOD) or by computing the reduced propagator and its eigendecomposition (DMD).

\begin{figure*}[t]
    \centering
    \includegraphics[width=\linewidth]{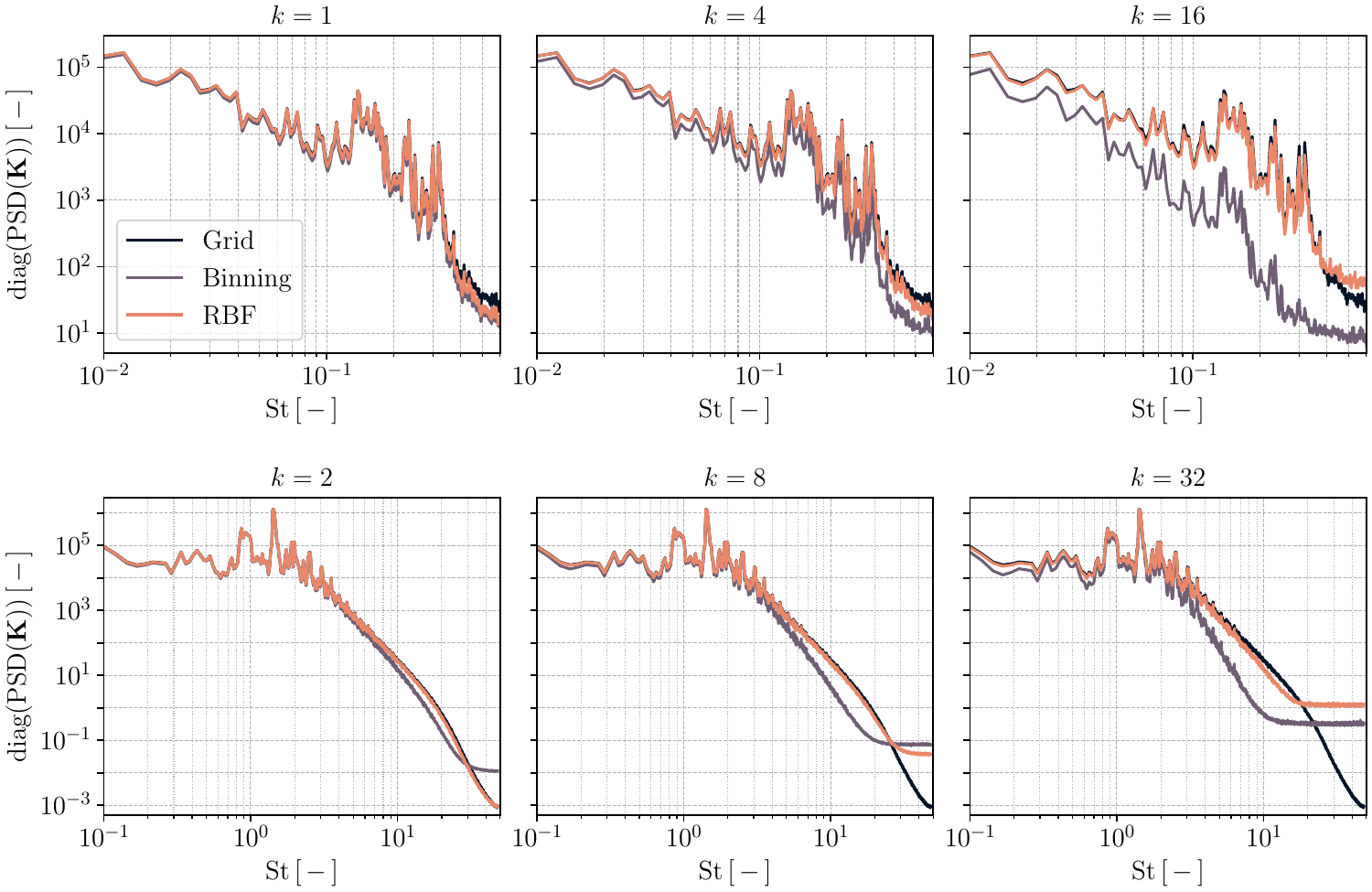}
    \caption{Diagonal of the power spectral density of the temporal correlation matrix $\mathbf{K}$ over the Strouhal number for the jet (top) and the airfoil (bottom). The power spectral density is computed using Welch's method with Hanning windows, \SI{75}{\percent} overlap and \num{500} (Jet) and \num{4000} (Airfoil) samples per segment. The columns correspond to increasing downsampling factors according to the title of each subpanel. The legend is shared across all subpanels and in regions of strong overlap, the curves match almost completely.}
    \label{fig:4_3_Diagonal_FFT_K}
\end{figure*}

\subsection{The correlation matrix}
\label{sec:4_1}

As a first estimate of how each method performs, we analyze the normalized error in the temporal correlation matrix $\mathbf{K}$, taken as: 
\begin{align}
    \label{eq:k_err}
    \vert\vert \mathbf{K} \vert\vert_\text{err} = \frac{\vert\vert \mathbf{K}_\text{exact} - \mathbf{K} \vert\vert_2}{\vert\vert \mathbf{K}_\text{exact}\vert\vert_2}\,,
\end{align} where $\mathbf{K}$ is the correlation matrix from either binning or the RBF-based inner product and the subscript 'exact' refers to the gridded data from the PIV and LES.

These errors over the downsampling factor $k$ are visualized in Figure~\ref{fig:4_1_K_convergence} for the jet and airfoil on the left- and right-hand side. The RBF error is consistently smaller than the binning one, reaching values below \SI{6}{\percent} for the jet and below \SI{1}{\percent} for the airfoil. The gap between the two methods is almost constant for larger $k$, but as the downsampling decreases, the gap narrows for the jet while it widens for the airfoil. We attribute this to the nature of the data: The PIV data are contaminated and biased by reflections near the wall. The RBFs partially filter these errors and thus never fully match the correlation matrix of the PIV. In contrast, the LES is noise-free and resolves more scales which is why more data also results in higher accuracy for the RBFs. The same is true for binning, although the benefits at lower $k$ are smaller.

The dashed circles in Figure~\ref{fig:4_1_K_convergence} indicate the three downsampling factors mentioned in section~\ref{sec:3}, and we analyze only these in the following to reduce the number of subpanels in the subsequent figures. For these three conditions, Figure~\ref{fig:4_2_K_matrices} shows the correlation matrix zoomed in on the first 200 timesteps. The left-hand side shows the jet and the right-hand side the airfoil results, while the rows respectively correspond to the gridded, binning, and RBF data. The title of the columns indicates the different downsampling factors. From the gridded results, the difference between the datasets is clearly visible. The jet has a coarse temporal resolution, resulting in small ``structures'' in the correlation matrix. In contrast, the airfoil has a fine temporal resolution, resulting in a visually much smoother matrix in the focused region. If the matrix is viewed with more timesteps, the airfoil case likewise shows a similarly ``complex'' pattern as the jet, albeit at much higher resolution.

For the smallest downsampling factor, both binning and the RBFs reproduce the respective patterns of the gridded data well. Then, as the downsampling factor increases, we observe two things. First, the magnitude of the correlation matrix decreases, which we attribute to the spatial low-pass filtering of both methods. Large fluctuations in space are either not present in the data or averaged out, leading to a smaller velocity magnitude and thus, a dampened correlation matrix. Second, the fine-scale structures in the matrix are smoothed out. This happens most severely for binning of the jet due to the coarse temporal resolution. This apparent smoothing is stronger for binning since it is a straightforward low-pass filter, while the regression properties of the RBFs mitigate spatial smoothing to some extent. These results also agree with the error curves in Figure~\ref{fig:4_1_K_convergence}, which are linked to a decreasing magnitude and thus, a larger error.

The correlation matrices of the Towne SPOD show similar patterns, which is why they are omitted. The RBFs also consistently reach a lower error in the magnitude of the correlation matrices across all frequencies, which is again linked to the low-pass filtering and attenuation of binning. Low-pass filtering in space evidently dampens the Fourier amplitudes, and hence the correlation matrix $\mathbf{M}$.

\begin{figure*}[tb]
    \centering
    \includegraphics[width=\linewidth]{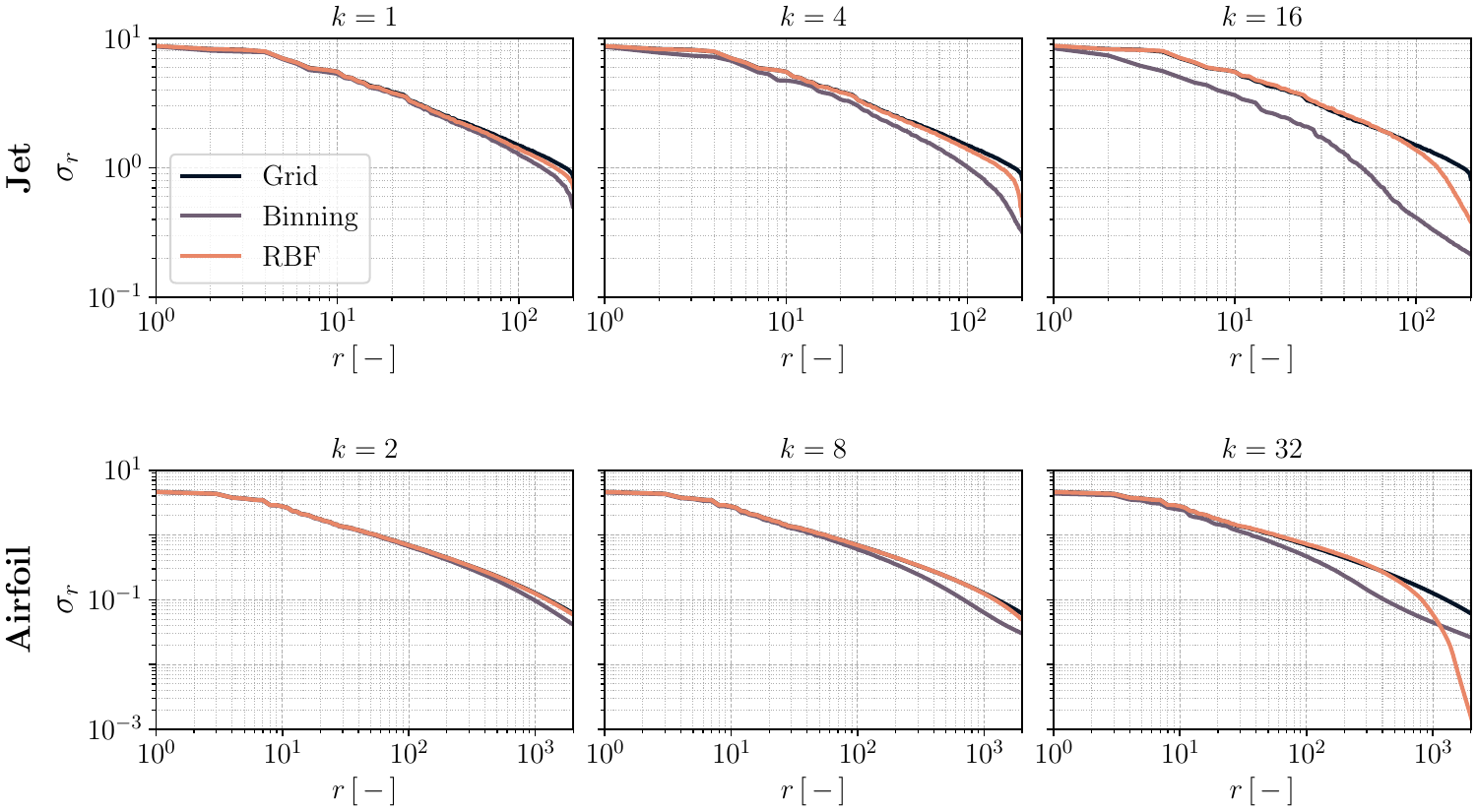}
    \caption{Modal amplitudes from the POD over the mode number $r$ for the jet (top) and the airfoil (bottom). The columns correspond to increasing downsampling factors according to the title of each subpanel. Only the first 200 and 2000 modes are shown for the jet and the airfoil, respectively. The legend is shared across all subpanels.}
    \label{fig:4_4_Sigmas_POD}
\end{figure*}

To assess the meshless mPOD and Sieber SPOD, we compare the spectral content of the correlation matrix $\hat{\mathbf{K}}$ by computing its power spectral density (PSD). Figure~\ref{fig:4_3_Diagonal_FFT_K} displays the diagonal of this matrix for the jet (top) and the airfoil (bottom). For visualization purposes, we use Welch's method with Hanning windows, \SI{75}{\percent} overlap, and 500 (Jet) and 4000 (Airfoil) samples per segment; the PSD without windowing shows the same trend.

For the jet, both methods yield similar results for $k=1$. The curves overlap almost completely and are indistinguishable except for the highest frequencies close to the Nyquist limit at $\Strouhal = 0.5$. The wide spectrum is clearly visible, with pronounced peaks up to $\Strouhal = 0.3$. For $k=4$, binning yields a slightly attenuated spectrum, particularly for frequencies above $\Strouhal = 0.1$. This attenuation is further amplified for $k=16$: The entire spectrum is dampened, the peak around $\Strouhal=0.2$ is barely visible, and the footprint of the Kelvin-Helmholtz rollers in the shear layer at $\Strouhal = 0.3$ is not visible. In contrast, the RBFs yield an almost perfect spectrum for $k=4$ and show only minor differences for $k=16$. Significant peaks are still present at all frequencies, and all relevant features are recovered.

We expect these observations to also affect the mPOD and Sieber SPOD. Binning cannot recover features at high Strouhal numbers, and we can hardly expect any filtering operation on the correlation matrix to do so. The same happens for the mPOD: what is lost in space and time cannot be recovered, no matter the decomposition. Since the RBFs yield a better spectrum of $\mathbf{K}$, we expect the meshless mPOD and Sieber SPOD to work equally well as in the gridded case.

In contrast, the airfoil case shows only minor differences in the spectrum in the relevant frequencies, even for $k=32$. A peak at $\Strouhal \approx 1.47$ identifies the frequency of the laminar separation bubble \cite{Taira2017}, and the harmonic peaks up to $\Strouhal = 3.5$ are clearly visible across all cases. Higher frequencies tell a different story. For both methods, the spectrum flattens and resembles white noise. For binning, this transition toward white noise happens at lower frequencies and is preceded by a dampened amplitude. Conversely, the RBFs follow the grid data almost perfectly until they drop off to white noise, the earliest frequency being $\Strouhal = 10$ for $k = 32$. We link this different behavior of the RBFs to the different datasets: Compared to the jet, the amplitude range of the PSD is twice as large. For the jet, almost all frequencies are recovered well since they contain significant (spatial) features, such as vortices, while the high frequencies of the airfoil show no significant features and have small amplitude. Because of this, we expect both the mPOD and the Sieber SPOD to work well for the dominant features of both binning and RBF.

\subsection{The modal amplitudes}
\label{sec:4_2}

\begin{figure*}[tb]
    \centering
    \includegraphics[width=\linewidth]{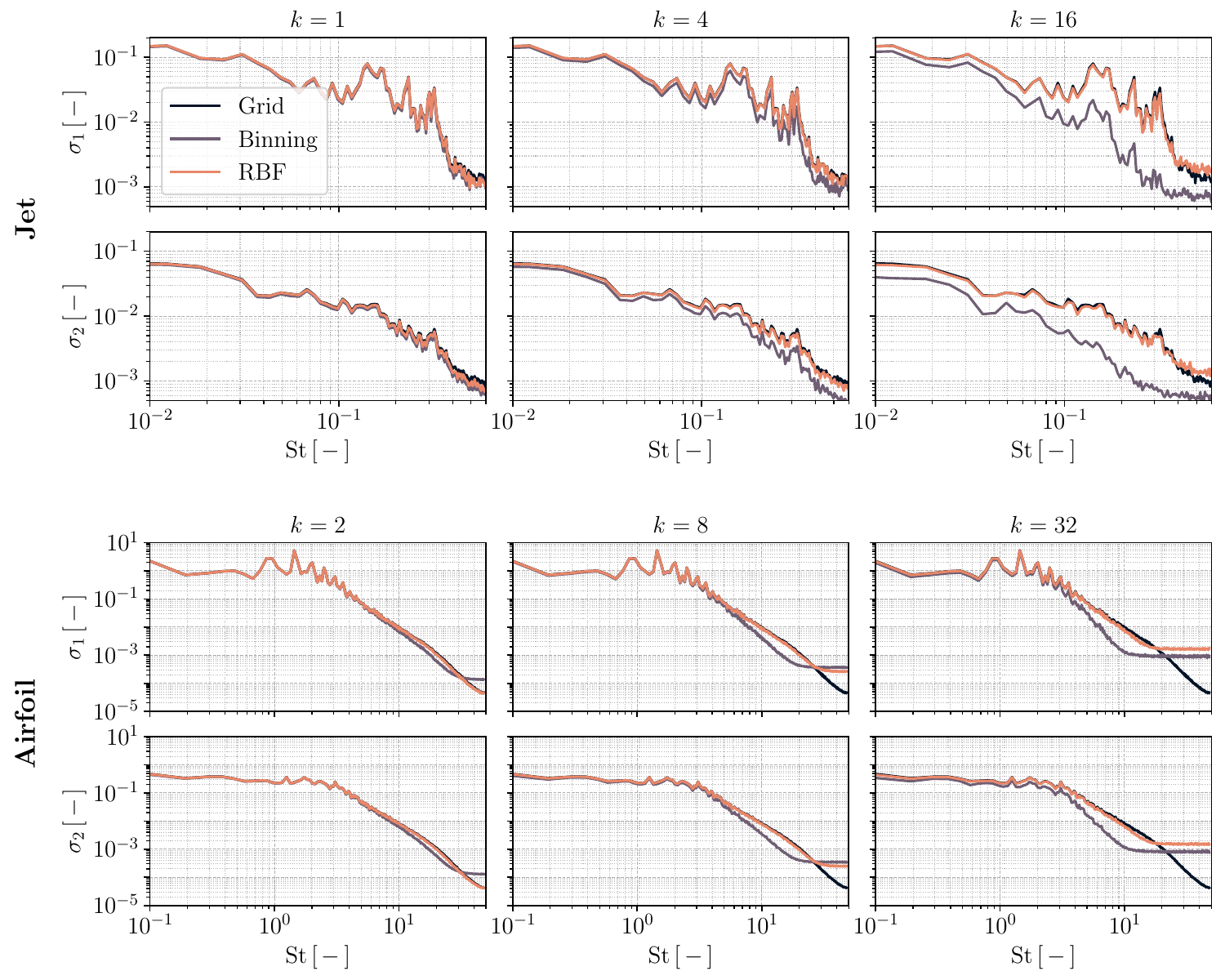}
    \caption{First and second modal amplitude of the Towne SPOD over the Strouhal number for the jet (row 1 and 2) and the airfoil (row 3 and 4). The columns correspond to increasing downsampling factors according to the title of each subpanel. The range of frequencies matches Fig. \ref{fig:4_3_Diagonal_FFT_K}, and the legend is shared across all subpanels.}
    \label{fig:4_5_Sigmas_SPOD_t}
\end{figure*}

Any error in the correlation matrices $\mathbf{K}$ and $\mathbf{M}$ will also influence their eigendecomposition. Since the eigenvectors are orthonormal, we expect a different magnitude of the matrix to affect only the modal amplitudes, whereas a different structure of the matrix can affect both the amplitudes and the eigenvectors. To investigate the first hypothesis, Figure~\ref{fig:4_4_Sigmas_POD} shows the modal amplitudes of the eigendecomposition of $\mathbf{K}$ for the jet (top row) and the airfoil (bottom row), while the columns correspond to different downsampling factors according to the title of each subpanel. For better comparison, we only show the first 200 and 2000 modes for the jet and airfoil. 

For the jet, both binning and RBF match the result of the gridded POD well for $k=1$. Modes above 100 are slightly attenuated, but the curves show the same trends. As the downsampling increases, the modal amplitudes are dampened further, albeit stronger for binning than for RBF. For $k=16$, already the third mode of binning has only \SI{75}{\percent} of the grid amplitude, and this gap is almost constant over a wide range of modes. In contrast, the RBFs show similar dampening only for modes above 100, albeit with a steeper drop-off than binning. The airfoil curves show similar trends, although less pronounced. For $k=2$, slight attenuation occurs only above mode \num{1000}, and even at $k=32$, binning captures the first 30 modes well. As before, the RBFs match the grid curve better, but then have a sudden, steep drop in the amplitude after approximately 500 modes for $k=32$. These differences between binning and RBF confirm the hypothesis about the correlation matrix: A dampened correlation matrix leads to dampened modal amplitudes.

One consistent issue is the steep drop-off in the modal amplitudes of the RBFs at high mode numbers. Larger eigenvalues beyond the range shown here are multiple orders of magnitude smaller. We attribute this to the numerical stability of the solver since a Sieber SPOD with a cut-off frequency at $\Strouhal=25$ (airfoil) already removes this sharp drop-off. The mPOD shows similar results. Any range of reasonable parameters results in modal amplitudes matching the gridded result much better than binning. The other observations remain, which is why we omit to show the modal amplitudes of these POD variants.

The decomposition of the correlation matrix $\mathbf{M}$ for the Towne SPOD yields multiple modal amplitudes at each frequency. Figure~\ref{fig:4_5_Sigmas_SPOD_t} depicts the two largest modal amplitudes over the Strouhal number, where the first and second rows correspond to the jet and the third and fourth to the airfoil. The columns again indicate the downsampling factor $k$, increasing from left to right.

\begin{figure*}[tb]
    \centering
    \includegraphics[width=\linewidth]{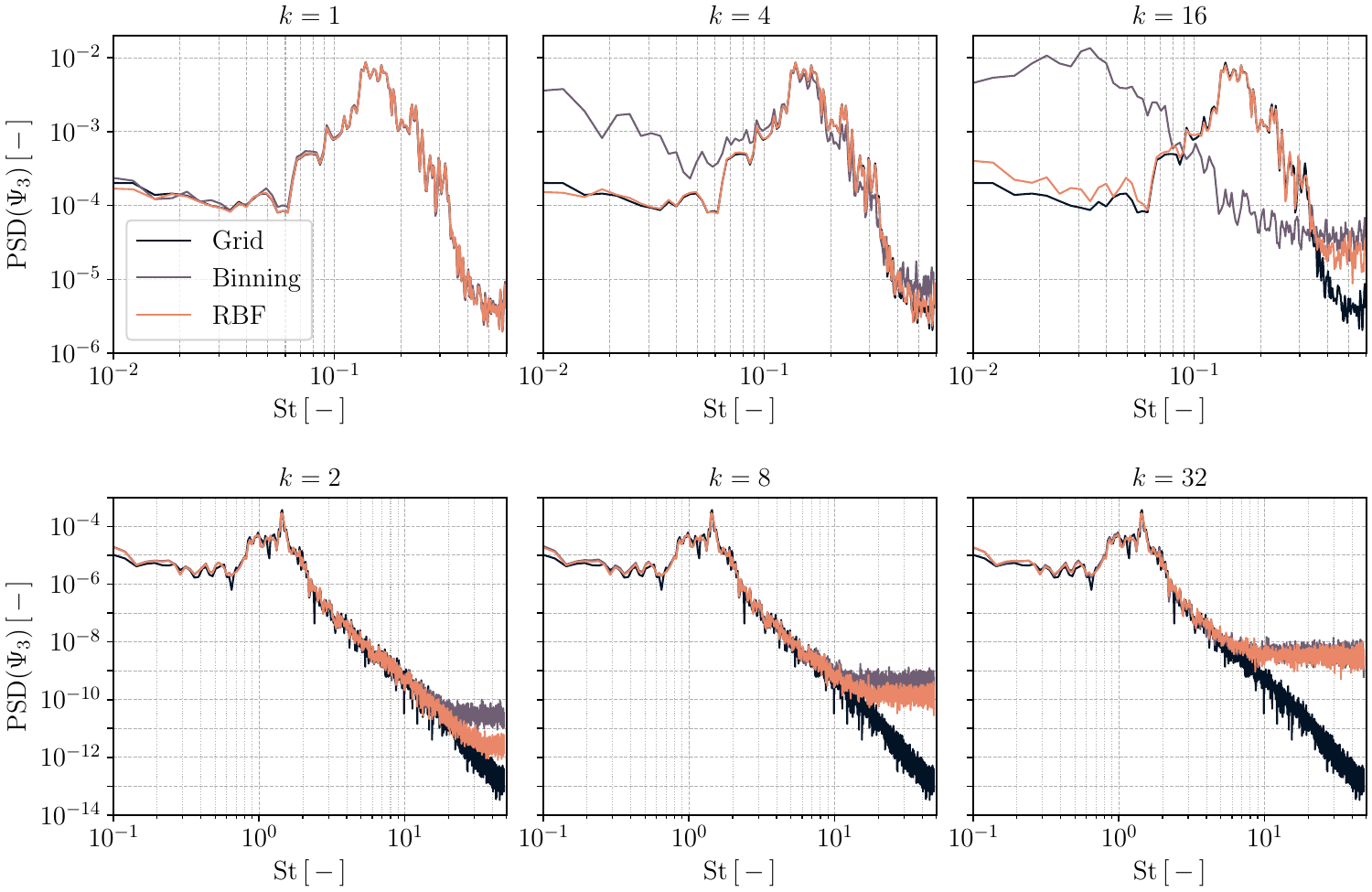}
    \caption{Power spectral density of the third POD mode over the Strouhal number for the jet (top) and the airfoil (bottom). The power spectral density is computed using Welch's method with Hanning windows, \SI{75}{\percent} overlap and \num{500} (Jet) and \num{4000} (Airfoil) samples per segment. The columns correspond to increasing downsampling factors according to the title of each subpanel. The legend is shared across all subpanels and in regions of strong overlap, the curves match almost completely.}
    \label{fig:4_6_Psi_POD}
\end{figure*}

For the jet test case, we obtain similar results as for the spectral analysis of the temporal correlation matrix. At small $k$, both methods yield more or less the same spectrum for the first and second modes. Then, as $k$ increases, binning attenuates the spectrum. For $k=4$, the differences are minor, but for $k=16$, the spectrum is flattened severely at high Strouhal numbers. However, compared to the spectral analysis of $\mathbf{K}$ (c.f. figure \ref{fig:4_3_Diagonal_FFT_K}), a small peak at $\Strouhal=0.3$ can still be identified for binning. This difference is expected, since the Towne SPOD is a more sophisticated analysis than a simple PSD of $\mathbf{K}$. Conversely, the RBFs almost perfectly recover the whole spectrum, with only some slight differences at high frequencies beyond $\Strouhal = 0.4$. This behavior is consistent across all mode numbers: Binning gives dampened spectra while the RBFs match almost perfectly.

The airfoil results agree with the previous analysis. In both the first and the second modes, the modal amplitudes of binning are slightly dampened. The higher $k$, the earlier this attenuation happens: at $\Strouhal = 3$ in the sparsest case. Afterward, the spectrum flattens to white noise with the cut-off frequency again decreasing with increasing $k$. In comparison, the RBFs either completely follow the spectrum $(k=1)$ or follow it up to a cut-off frequency and then transition with a sharp kink toward white noise. This cut-off frequency is above $\Strouhal=10$ for both modes and always higher for RBF than for binning. Moreover, it is almost constant for all modes. In summary, the SPOD results agree well with the spectral analysis of the temporal correlation matrix: The RBFs follow the spectrum of the grid more faithfully and up to a higher Strouhal number.

\subsection{The temporal basis}
\label{sec:4_3}

For the POD, mPOD, and Sieber SPOD, the eigendecomposition of $\mathbf{K}$ gives the temporal structures. Figure~\ref{fig:4_6_Psi_POD} shows the PSD of the third POD mode for the jet (top) and the airfoil (bottom), while the columns again correspond to an increasing $k$ from left to right, as indicated by the titles. For better comparison, the spectra are again visualized with the same Welch settings as in Figure~\ref{fig:4_3_Diagonal_FFT_K}.

The chosen mode of the jet contains a pronounced peak around $\Strouhal = 0.15$, and both methods recover it equally well. Compared to the Nyquist frequency, this frequency is sufficiently small such that both methods recover it well for $k=1$. However, as the downsampling increases, binning shows spectral mixing: the low-frequency components have a higher amplitude than the gridded and RBF versions. For $k=4$, the peak at $\Strouhal = 0.15$ can still be identified, but for $k=16$, it is completely gone. The RBFs also show signs of spectral mixing at $k=16$, with a slightly increased low and high-frequency component, but not as severe as binning. The third mode still clearly matches the gridded case.

The temporal structures of the mPOD and Sieber SPOD confirm the observations of the temporal correlation matrix, which is why they are omitted. At high $k$, high frequencies are diminished, which also translates into both POD variants. At $k=16$, no filter for the SPOD and neither the mPOD can recover the features of the Kelvin Helmholtz instability sufficiently well. What is lost cannot be recovered, no matter how sophisticated the decomposition is.

For the airfoil, spectral mixing is not present in the low-frequency content of the first modes. The peak at $\Strouhal \approx 1.5$ and the general shape of the spectrum are recovered almost perfectly for both methods and all downsampling factors. The only issues arise at high frequencies above $\Strouhal = 10$. The spectra of RBF and binning are again white noise, with a decreasing cut-off frequency for increasing $k$. We observe this white noise band in all modes; the cut-off frequency slightly varies but always has the same order of magnitude. Compared to the white noise behavior of $\mathbf{K}$ and the Towne SPOD (c.f. Figs.~\ref{fig:4_3_Diagonal_FFT_K} and \ref{fig:4_5_Sigmas_SPOD_t}), the cut-off frequency is smaller. The RBFs still perform slightly better than binning, but the differences are minor: significant features are located low small frequencies compared to the Nyquist limit because of the excellent temporal sampling. This POD mode, and low POD modes in general, are identified well.

However, this broadband, white noise behavior only affects the classical POD since it is not constrained or penalized in the frequency content of each mode. For the mPOD, high frequencies are filtered naturally due to the different filter banks; modes only contain white noise if they are from the filter band above $\Strouhal = 10$. Yet, since no relevant features are present at this frequency, the results of the mPOD are barely influenced by this. Neither are the results of the Sieber SPOD. For different finite impulse response filters, we observed the same trend: If the white noise band is present after a certain frequency, a filter with a cut-off frequency at this frequency always removes the band without affecting the dominant features of the mode. Hover these are minor details overall. The temporal resolution of the airfoil dataset is sufficiently large so that no relevant feature is drowned noise. In the jet, this behavior is not visible since almost all frequencies have relevant content, and the PSD spans fewer orders of magnitude.

\subsection{The spatial basis}
\label{sec:4_4}

\begin{figure*}[t]
    \centering
    \includegraphics[width=.91\linewidth]{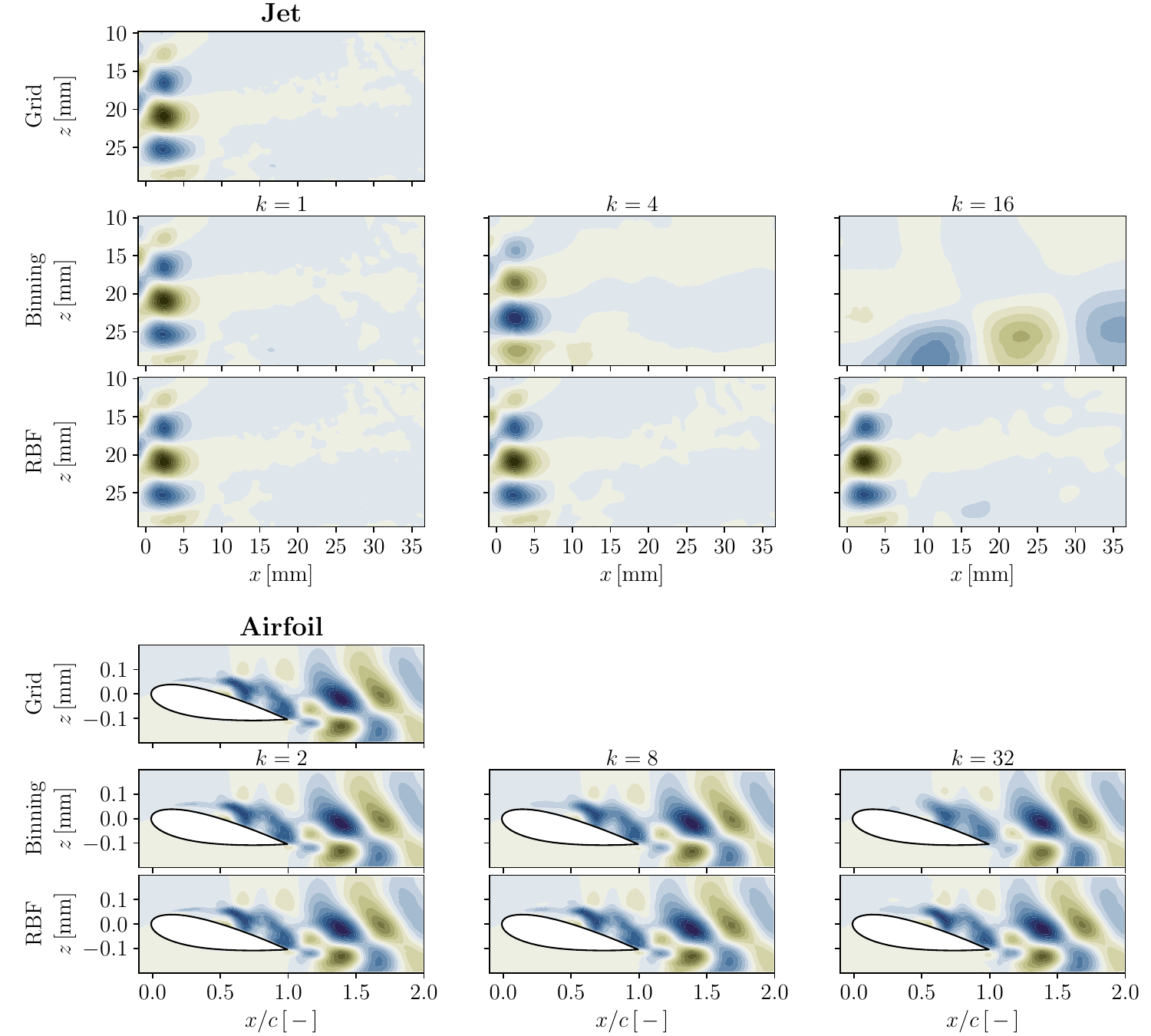}
    \caption{Spatial structures of the third POD mode for the jet (top part) and the airfoil (bottom part). The top left subpanel each shows the ground truth from the gridded data and the rows correspond to the method labeled on the left. The columns correspond to increasing downsampling factors according to the title of each subpanel and the range of contours is shared across the panels of each case. For the airfoil, the aspect ratio is slightly increased for better visualization.}
    \label{fig:4_7_Phi_POD}
\end{figure*}

We close the discussion of the different PODs by analyzing the resulting spatial structures. To this end, Figure~\ref{fig:4_7_Phi_POD} shows the third spatial structures, corresponding to the temporal mode of Figure~\ref{fig:4_6_Psi_POD}. The upper and lower part of the figure show the modes of the jet and the airfoil with the top left subpanel showing the gridded ground truth and the second and third row the results from binning and the RBFs. The columns again correspond to different downsampling factors $k$.

For the jet, at $k=1$, both binning and RBF show excellent agreement with gridded data. The Kelvin-Helmholtz rollers of the jet are clearly visible and look almost the same for all three methods. When fewer data points are available at $k=4$, the RBFs remain largely unchanged, while we can observe two things for binning. First, small structures of the developing wall jet at the bottom become visible. This agrees well with the observations in Figure~\ref{fig:4_6_Psi_POD}, where spectral mixing produced a temporal structure with multiple scales in time, leading to a spatial structure with multiple scales in space. The spectral mixing also likely causes the second observation: the Kelvin-Helmholtz rollers are out of phase compared to before. Space-only PODs always need at least two modes in quadrature to represent vortex shedding, but the respective spatial and temporal frequencies can also be present over a broader range of frequencies. The classic POD is affected most by this, while the slightly constrained Sieber and mPOD can somewhat mitigate this effect. Finally, at $k=16$, the Kelvin-Helmholtz instability is barely visible, and the mode is dominated by the wall jet for binning, again agreeing with the temporal structures from before. Intermediate downsampling factors between the three displayed ones show a gradual shift between these two extremes for the mode. In contrast, the RBFs only show an additional minor near-wall structure---the Kelvin-Helmholtz rollers are still clearly visible with unchanged frequency or amplitude. In conclusion, if the temporal structures do not match, we can hardly expect the spatial structures to do so since they are computed from the projection of the data onto the temporal modes.

\begin{figure*}[t]
    \centering
    \includegraphics[width=\linewidth]{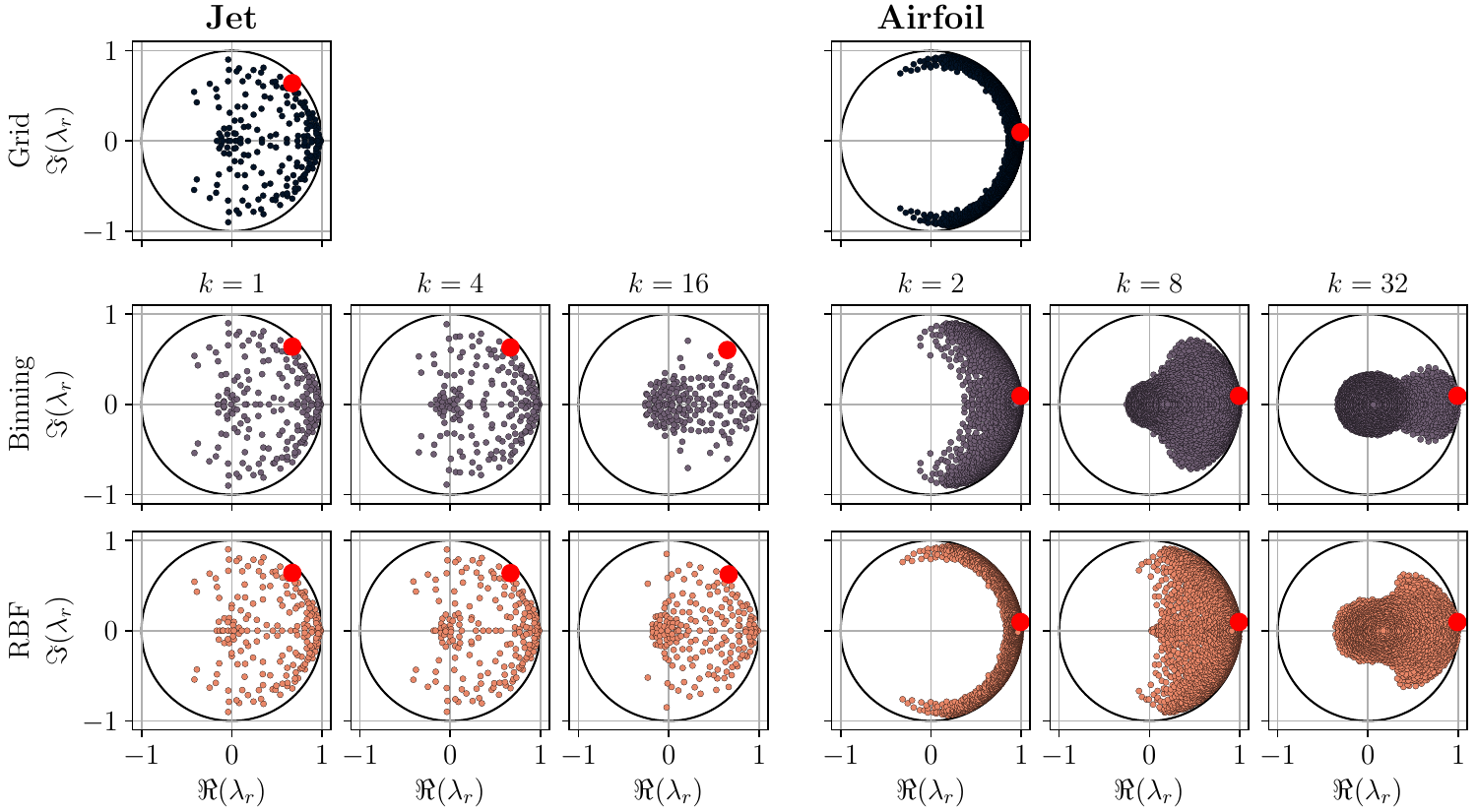}
    \caption{Eigenvalue distribution of the reduced propagator from \eqref{eq:reduced_propagator} in the complex plane for the jet (left) and the airfoil (right). The top left of each side shows the gridded case with 200 and 2000 POD modes for the jet and the airfoil. The center and bottom row correspond to binning and RBFs and each column contains one downsampling factor according to the title of the subpanel. The red dots indicate the eigenvalue of a temporal structure analyzed in Fig.~\ref{fig:4_9_DMD_Phi}.}
    \label{fig:4_8_DMD_Lambda}
\end{figure*}

Conversely, for the airfoil, the temporal structures match well for all $k$ and both methods, safe for some broadband content in the spectrum. The spatial structures in the lower half of Figure~\ref{fig:4_7_Phi_POD} clearly all show similar vortex shedding in the wake and a dominant structure over the latter half of the chord. While this is a qualitatively good agreement, this test case shows a different error of any of the PODs. The spatial basis is computed by projecting the data (or weights) onto the temporal structures. Any low-pass filtering in the data thus also affects the spatial structures, even if the temporal structures and modal amplitudes match perfectly. This is clearly visible in the fine-scale structures between the leading edge and half the chord length: a thin structure is visible for the gridded case, and also binning and RBF at $k=2$. This is associated with the edge of the laminar separation bubble, and the resulting structures in the latter half of the chord are associated with the transitioning shear layer. With increasing $k$, these small structures start to be smoothed: at $k=32$, the structures of separation are gone, and the ones of the transitioning shear layer are attenuated. In contrast, the structures for the RBFs only have a slightly diminished amplitude and are still clearly visible. We found this behavior to be consistent across all test cases and all variants of the POD: If the correlation matrices $\mathbf{K}$ or $\mathbf{M}$ and their eigendecomposition match the gridded case, then so do their spatial structures, save for some spatial smoothing depending on the data mapping method. Since the RBFs are more sophisticated and can better fill gaps than binning, they yield more accurate spatial structures.

In summary, the results of the different PODs clearly show the limitations of both binning and RBF. If the spectrum of the data contains significant features up to the acquisition frequency, as it does for the jet, good spatial mapping is required. If the resolution is too coarse, binning exhibits increased spectral mixing and thus produces poor spatial structures. The RBFs are more sophisticated and can thus recover the relevant spatial and temporal structures, even in challenging conditions. If the relevant frequencies are significantly smaller than the Nyquist limit, we can afford a simpler spatial mapping since the high temporal resolution helps to recover the correct temporal and spatial modes. However, even if the modal amplitudes and temporal structures are perfect, the spatial structures can still be oversmoothed since they are computed by projecting in space. Therefore, RBFs remain superior to binning since they result in more accurate spatial structures across all data densities.

\begin{figure*}[t]
    \centering
    \includegraphics[width=.91\linewidth]{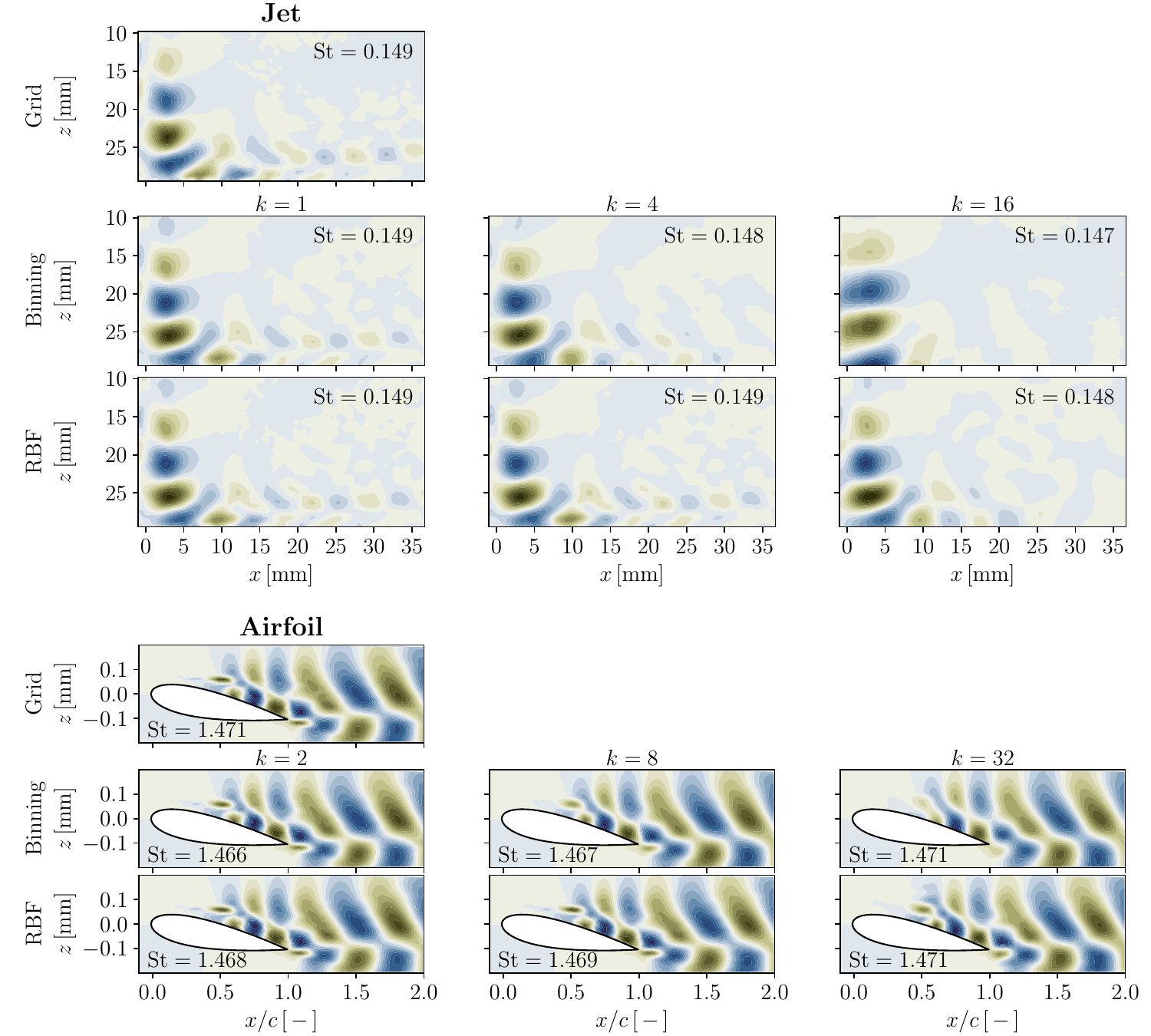}
    \caption{Spatial structures of a dominant frequency marked with a red dot in Fig.~\ref{fig:4_8_DMD_Lambda} for the jet (top) and the airfoil (bottom). The top left subpanel each shows the ground truth from the gridded data and the rows correspond to the method labeled on the left. The columns correspond to increasing downsampling factors according to the title of each subpanel and the range of contours is shared across the panels of each case. In each subpanel, the frequency of each mode is also given. For the airfoil, the aspect ratio is slightly increased for better visualization.}
    \label{fig:4_9_DMD_Phi}
\end{figure*}

\subsection{The Dynamic Mode Decomposition}
\label{sec:4_5}

The DMD uses the temporal structures and modal amplitudes from the POD to compute the reduced propagator. Errors in these two quantities will thus also affect the DMD. Figure~\ref{fig:4_8_DMD_Lambda} shows the resulting eigenvalues $\boldsymbol{\lambda}$ of the propagator in the complex plane, for the jet on the left and for the airfoil on the right. The top left panel shows the grid result, while the second and third rows again correspond to binning and RBF, with the columns indicating different downsampling factors $k$. The red dot in each figure corresponds to a frequency whose spatial structure is analyzed further below. 

For the jet at low $k$, both methods recover well the distribution of modes from the gridded data. Since the DMD is computed through the POD, only low-frequency modes with large real parts lie close to the unit circle. High-frequency modes are severely dampened. As $k$ increases, the amplitude of most eigenvalues decreases, and only few modes remain pronounced and dominant. This occurs more strongly for binning where at $k=16$, most eigenvalues are close to the origin and thus contribute only marginally to the decomposition. For the RBFs, more modes lie closer to the unit circle, and their magnitude is also larger for larger values of $k$. These observations agree well with the modal amplitudes in Figure~\ref{fig:4_4_Sigmas_POD}. For the jet, the higher modes at $k=16$ quickly have an attenuated amplitude for binning, while the ones of the RBF remain close to the gridded ones up to mode number 100. These dampened amplitudes evidently also affect the DMD eigenvalues. They are most dominant at low frequencies, corresponding well to the spectra of the correlation matrix $\mathbf{K}$ in Figure~\ref{fig:4_3_Diagonal_FFT_K}, where high frequencies are strongly dampened for binning but not for the RBFs. These properties are inherited via the POD of $\mathbf{K}$ to the eigenvalues of the DMD propagator.

The airfoil observations are similar. For the gridded case, more modes are close to the unit circle, even modes with a high Strouhal number. For $k=2$, binning already has many modes with smaller magnitudes, while the RBFs reproduce the apparent pattern well. As $k$ increases, many modes are starting to be far from the unit circle. When building a reduced order model with binning, only modes of relatively low frequency will be present for longer than a few timesteps. The RBFs are affected less severely, although many modes still have a small amplitude. Finally, at the highest downsampling factor, only few modes at low frequency are close to the unit circle for both binning and the RBFs. Even though the RBFs match the POD amplitudes of the gridded case for a larger number of modes, this appears to be an insufficient condition to have high-frequency eigenvalues close to the unit circle. Spectral analysis of these correctly matched POD modes showed they indeed mostly contained small frequencies while high-frequency content is captured by neither binning nor the RBF POD and hence, also not in the DMD.

Finally, regarding the spatial structures of the DMD, Figure~\ref{fig:4_9_DMD_Phi} shows the real part of spatial modes marked by the large red dots in Figure~\ref{fig:4_8_DMD_Lambda}. These are around $\Strouhal = 0.15$ for the jet and $\Strouhal=1.47$ for the airfoil; both are dominant frequencies in the spectra of the temporal correlation matrix in Figure~\ref{fig:4_3_Diagonal_FFT_K}. As before for the POD, the upper and lower part show the jet and the airfoil, the rows the method according to the label, and the columns the downsampling factor $k$.

For the jet, we can make similar observations as for the spatial structures of the POD. At the lowest $k$, fine-scale structures are recovered well. In the developing wall jet, the vortices are clearly visible for both binning and RBF. A frequency shift is observed compared to the gridded case, but this is simply a matter of also considering the imaginary part. As $k$ increases, these structures are diminished, and larger structures in the free jet are smeared out for binning. Since the DMD produces only one frequency per mode, there is no spectral mixing; instead, small structures near the wall are smoothed out. This effect is not as extreme for the RBFs, as fine structures near the wall are still visible at $k=16$. The identified frequency of this mode varies little, with a maximum deviation of $\Delta \Strouhal = 0.002$ for binning at $k=16$. Other dominant structures are found equally well, with similar, small differences in frequency.

For the airfoil, the same observations hold. At low $k$, all three methods are almost indistinguishable, and the small structures of the separation bubble are recovered well. As $k$ increases, both methods start to suffer, with binning suffering more than the RBFs since the spatial low-pass filtering of the method translates into smoothed spatial structures. Also, the difference in the detected frequency is small across both methods and all values of the downsampling factor $k$.

In summary, since the DMD is computed through the POD, it inherits some of the faults: high frequencies are diminished in the POD spectra for larger values of $k$ and thus, also in the DMD. The spatial structures likewise are affected by the spatial smoothing of the data mapping method, leading to smoothed structures. Spectral mixing is alleviated since the DMD gives only one frequency per mode

\section{Conclusion and Perspectives}
\label{sec:5}

We propose a meshless framework to compute data-driven modal decompositions of scattered data. The method uses physics-constrained radial basis functions (RBFs) to obtain an analytical expression of the data at every time instance. This expression is used to compute the inner products required for every modal decomposition, namely the proper orthogonal decomposition (POD), the dynamic mode decomposition (DMD), the two spectral PODs (SPODs), and multi-scale POD (mPOD). All inner products in space of the data are computed through cheap Monte Carlo sampling of the basis functions, greatly reducing computational cost and allowing reuse of expressions between the decompositions. The approach results in super-resolution of all spatial structures, meaning they can be expressed in an arbitrary set of points.

We compare our proposed method with binning, a simple moving average over a discrete set of points that approximates the spatial inner product with summations. We evaluate both methods on two different test cases: an impinging jet on a plate from particle image velocimetry (PIV) and a flow past a transitional airfoil from large eddy simulations (LES). Both cases feature a wide range of scales. Yet, the temporal and spatial resolution of the PIV is, typical for experiments, coarse, while both are finer for the LES. For both datasets, we generate scattered data with different spatial downsamplings to compare how binning and RBF-based inner products handle varying data densities.

For binning, the temporal correlation matrix appears smoothed and dampened, and its spectra show diminished higher frequencies. Instead, the RBFs almost perfectly match the PIV and LES in the relevant range of frequencies, even when reducing the original data density by factors of 16 or larger. These effects are much stronger for the jet, where the temporal resolution is limited; for the airfoil, the high temporal resolution is able to mitigate the loss of spatial information.

Nevertheless, the performance of the meshless decomposition framework is inherently tied to the quality of the RBF regression and thus to the spatial sampling density. The results presented here are obtained under conditions where the seeding density is sufficiently high for the RBF approximation to resolve the relevant spatial gradients and thereby yield more accurate inner products, spectra, and modal shapes. Although the RBF approach is known to systematically outperform binning as the seeding density decreases (\cite{Ratz2024}), a reduction in seeding inevitably reduces its advantage.

These observations translate from the eigenvalue decompositions of the correlation matrix for the POD, mPOD, and SPOD \citep{Sieber2016} into the modal amplitudes and temporal structures. Also for the Fourier-based SPOD \citep{Towne2018}, the RBF-based inner products outperform binning and give more accurate spectra for both the PIV and the LES. Binning always attenuates higher frequency content if temporal and spatial resolution are insufficient. In these conditions, we also observe increased spectral mixing in the POD modes for binning while the RBFs remain largely unaffected. An mPOD with filter banks at high frequencies or a Sieber SPOD with large filter frequencies already resolves this issue.

The spatial structures are computed from the projection of the data onto the temporal matrix. Therefore, if the temporal structures are affected by the spectral mixing, then so are the spatial structures. Also, since the data incorporates any low-pass filtering of the method, this also translates into the spatial structures. Thus, even if a perfect amplitude and temporal structure are computed, binning will always show low-pass filtering, which the RBFs do not since regression is a more sophisticated method. These observations are consistent across all of the mentioned POD variants.

The DMD also inherits these properties since it is computed from a truncated POD. If the data are downsampled, higher frequency content is attenuated in the temporal structures of the POD and also the eigenvalues of the reduced propagator, which are located further from the unit circle. This occurs more severely for high-frequency modes and more for binning than for the RBFs. The spatial structures are computed from the POD and thus suffer the same fate: If the temporal structure is poor, then so is the spatial one, and any spatial filtering in the data mapping method also translates into the spatial structure. Since the RBFs can better fill the gaps of scattered data, they naturally cope with this better.

The most natural path for future research is to also include temporal information. For binning, no off-the-shelf method exists, but 4D extensions with a Gaussian or binomial weight seem a feasible way to incorporate spatio-temporal smoothing. For the RBF regression, temporal information has been used by adding temporal super-resolution of Lagrangian tracks \citep{Gesemann2016} and iterative updating between tracks and snapshots \citep{Li2024}. An analytical expression of the data in time would also allow to compute the inner products in time when the data are projected onto the temporal modes.

\section*{Acknowledgements}
M.R. is funded by the Fonds de la Recherche Scientifique (F.R.S.-FNRS) through the FRIA grant No. FC57471. M.A.M.'s contributions are supported by the European Research Council (ERC) under the Horizon Europe programme (Starting Grant No. 101165479, RE-TWIST). The data are visualized using perceptually uniform colormaps to prevent distortion of the data and exclusion of readers with color-vision deficiencies \cite{Crameri2020, Crameri2023}.

\bibliographystyle{elsarticle-num-names}
\bibliography{bibliography}

@Article{Towne2018,
  author  = {Towne, Aaron and Schmidt, Oliver T. and Colonius, Tim},
  journal = {Journal of Fluid Mechanics},
  title   = {Spectral proper orthogonal decomposition and its relationship to dynamic mode decomposition and resolvent analysis},
  year    = {2018},
  pages   = {821–867},
  volume  = {847},
  doi     = {10.1017/jfm.2018.283},
}

@Article{Schmid2010,
  author  = {Schmid, Peter J.},
  journal = {Journal of Fluid Mechanics},
  title   = {Dynamic mode decomposition of numerical and experimental data},
  year    = {2010},
  pages   = {5–28},
  volume  = {656},
  doi     = {10.1017/S0022112010001217},
}

@Article{Tu2014,
  author    = {Tu, Jonathan H. and Rowley, Clarence W. and Luchtenburg, Dirk M. and Brunton, Steven L. and Kutz, J. Nathan},
  journal   = {Journal of Computational Dynamics},
  title     = {On dynamic mode decomposition: Theory and applications},
  year      = {2014},
  issn      = {2158-2505},
  number    = {2},
  pages     = {391--421},
  volume    = {1},
  doi       = {10.3934/jcd.2014.1.391},
  publisher = {American Institute of Mathematical Sciences (AIMS)},
}

@Article{Sieber2016,
  author  = {Sieber, Moritz and Paschereit, C. Oliver and Oberleithner, Kilian},
  journal = {Journal of Fluid Mechanics},
  title   = {Spectral proper orthogonal decomposition},
  year    = {2016},
  pages   = {798–828},
  volume  = {792},
  doi     = {10.1017/jfm.2016.103},
}

@Article{Mendez2020,
  author    = {Mendez, M. A. and Hess, D. and Watz, B. B. and Buchlin, J.-M.},
  journal   = {Measurement Science and Technology},
  title     = {Multiscale proper orthogonal decomposition ({mPOD}) of {TR-PIV} data—a case study on stationary and transient cylinder wake flows},
  year      = {2020},
  month     = {jun},
  number    = {9},
  pages     = {094014},
  volume    = {31},
  comment   = {mPOD application paper with the cylinder data},
  doi       = {10.1088/1361-6501/ab82be},
  publisher = {IOP Publishing},
}

@Article{Mendez2019,
  author  = {Mendez, M. A. and Balabane, M. and Buchlin, J.-M.},
  journal = {Journal of Fluid Mechanics},
  title   = {Multi-scale proper orthogonal decomposition of complex fluid flows},
  year    = {2019},
  pages   = {988–1036},
  volume  = {870},
  doi     = {10.1017/jfm.2019.212},
}

@Article{Hasselmann1988,
  author    = {Hasselmann, K.},
  journal   = {Journal of Geophysical Research: Atmospheres},
  title     = {PIPs and POPs: The reduction of complex dynamical systems using principal interaction and oscillation patterns},
  year      = {1988},
  issn      = {0148-0227},
  month     = sep,
  number    = {D9},
  pages     = {11015--11021},
  volume    = {93},
  doi       = {10.1029/jd093id09p11015},
  publisher = {American Geophysical Union (AGU)},
}

@Article{Storch1990,
  author    = {von Storch, Hans and Xu, Jinsong},
  journal   = {Climate Dynamics},
  title     = {Principal oscillation pattern analysis of the 30- to 60-day oscillation in the tropical troposphere: Part I: Definition of an index and its prediction},
  year      = {1990},
  issn      = {1432-0894},
  month     = sep,
  number    = {3},
  pages     = {175--190},
  volume    = {4},
  doi       = {10.1007/bf00209520},
  publisher = {Springer Science and Business Media LLC},
}

@article{Penland1996,
	title        = {A stochastic model of {IndoPacific} sea surface temperature anomalies},
	author       = {C{\'{e}}cile Penland},
	year         = 1996,
	month        = nov,
	journal      = {Physica D: Nonlinear Phenomena},
	publisher    = {Elsevier {BV}},
	volume       = 98,
	number       = {2-4},
	pages        = {534--558},
	doi          = {10.1016/0167-2789(96)00124-8}
}

@inbook{Mendez_2025_book2, 
    place={von Karman Institute for Fluid Dynamics},
    title={Fundamentals of Dimensionality Reduction}, 
    booktitle={Machine Learning for Fluid Dynamics}, 
    publisher={Cambridge University Press}, 
    author={Mendez, M.A. and Poletti, R. and Schena, L.}, 
    year={2025}, 
    pages={259-299}
}

@Article{Lumley1967,
  author  = {Lumley, John Leask},
  journal = {Atmospheric turbulence and radio wave propagation},
  title   = {The structure of inhomogeneous turbulent flows},
  year    = {1967},
  pages   = {166--178},
}

@Article{Sirovich1987,
  author  = {Lawrence Sirovich},
  journal = {Quarterly of Applied Mathematics},
  title   = {Turbulence and the dynamics of coherent structures. Part {I}: {C}oherent structures},
  year    = {1987},
  issn    = {0033569X, 15524485},
  number  = {3},
  pages   = {561--571},
  volume  = {45},
}

@Article{Welch1967,
  author   = {Welch, P.},
  journal  = {IEEE Transactions on Audio and Electroacoustics},
  title    = {The use of fast {F}ourier transform for the estimation of power spectra: A method based on time averaging over short, modified periodograms},
  year     = {1967},
  number   = {2},
  pages    = {70--73},
  volume   = {15},
  doi      = {10.1109/TAU.1967.1161901},
}

@InCollection{Lumley1981,
  author    = {J. L. Lumley},
  booktitle = {Transition and Turbulence},
  publisher = {Academic Press},
  title     = {Coherent {S}tructures in {T}urbulence},
  year      = {1981},
  editor    = {R. E. Meyer},
  isbn      = {978-0-12-493240-1},
  pages     = {215--242},
  comment   = {Seminal paper by Lumley on POD application},
  doi       = {10.1016/B978-0-12-493240-1.50017-X},
}

@Article{Rowley2017,
  author    = {Rowley, Clarence W. and Dawson, Scott T. M.},
  journal   = {Annual Review of Fluid Mechanics},
  title     = {Model {R}eduction for {F}low {A}nalysis and {C}ontrol},
  year      = {2017},
  issn      = {1545-4479},
  number    = {Volume 49, 2017},
  pages     = {387-417},
  volume    = {49},
  comment   = {Overview paper for POD applications},
  doi       = {10.1146/annurev-fluid-010816-060042},
  publisher = {Annual Reviews},
  type      = {Journal Article},
}

@Article{Berkooz1993,
  author    = {Berkooz, G. and Holmes, P. and Lumley, J. L.},
  journal   = {Annual Review of Fluid Mechanics},
  title     = {The {P}roper {O}rthogonal {D}ecomposition in the {A}nalysis of {T}urbulent {F}lows},
  year      = {1993},
  issn      = {1545-4479},
  number    = {Volume 25, 1993},
  pages     = {539-575},
  volume    = {25},
  comment   = {Historical review paper of the POD which introduces it as an integral instead of the usual matrix factorization},
  doi       = {10.1146/annurev.fl.25.010193.002543},
  publisher = {Annual Reviews},
  type      = {Journal Article},
}

@Article{Towne2023,
  author  = {Towne, Aaron and Dawson, Scott T. M. and Br\`{e}s, Guillaume A. and Lozano-Dur\'{a}n, Adri\'{a}n and Saxton-Fox, Theresa and Parthasarathy, Aadhy and Jones, Anya R. and Biler, Hulya and Yeh, Chi-An and Patel, Het D. and Taira, Kunihiko},
  journal = {AIAA Journal},
  title   = {A {D}atabase for {R}educed-{C}omplexity {M}odeling of {F}luid {F}lows},
  year    = {2023},
  number  = {7},
  pages   = {2867--2892},
  volume  = {61},
  comment = {Excellent dataset description for modal analysis},
  doi     = {10.2514/1.J062203},
}

@Article{Yeh2019,
  author  = {Yeh, Chi-An and Taira, Kunihiko},
  journal = {Journal of Fluid Mechanics},
  title   = {Resolvent-analysis-based design of airfoil separation control},
  year    = {2019},
  pages   = {572–610},
  volume  = {867},
  comment = {Dataset of the transitional airfoil},
  doi     = {10.1017/jfm.2019.163},
}

@Article{Feeny1998,
  author    = {Feeny, Brian F. and Kappagantu, Rajesh},
  journal   = {Journal of Sound and Vibration},
  title     = {On the physical interpretation of proper orthogonal modes in vibrations},
  year      = {1998},
  number    = {4},
  pages     = {607--616},
  volume    = {211},
  comment   = {Apply POD to simple vibration analysis of mass spring systems},
  doi       = {10.1006/jsvi.1997.1386},
  publisher = {Elsevier},
}

@InProceedings{Nonnenmacher2017,
  author        = {Marcel Nonnenmacher and Srinivas C. Turaga and Jakob H. Macke},
  booktitle     = {31st Conference on Neural Information Processing Systems},
  title         = {Extracting low-dimensional dynamics from multiple large-scale neural population recordings by learning to predict correlations},
  year          = {2017},
  comment       = {Use POD to get a low dim representation of neuroscience data},
  pages         = {1-11},
}

@Article{Shi2014,
  author    = {Shi, R. X. and Wang, J. F. and Xu, C. D. and Lai, S. J. and Yang, W. Z.},
  journal   = {Public Health},
  title     = {Spatiotemporal pattern of hand–foot–mouth disease in {C}hina: an analysis of empirical orthogonal functions},
  year      = {2014},
  issn      = {0033-3506},
  month     = apr,
  number    = {4},
  pages     = {367--375},
  volume    = {128},
  comment   = {Application of the POD to epidemiology},
  doi       = {10.1016/j.puhe.2014.01.005},
  publisher = {Elsevier BV},
}

@Book{Mendez2023,
  editor    = {Mendez, M. A. and Ianiro, A. and Noack, B. R. and Brunton, S. L.},
  publisher = {Cambridge University Press},
  title     = {Data-{D}riven {F}luid {M}echanics: {C}ombining {F}irst {P}rinciples and {M}achine {L}earning},
  year      = {2023},
  doi       = {10.1017/9781108896214},
  place     = {Cambridge},
}

@article{Monaghan1992,
  title={Smoothed particle hydrodynamics},
  author={Monaghan, J. J.},
  journal={Annual Review of Astronomy and Astrophysics},
  volume={30},
  number={1},
  pages={543--574},
  year={1992},
  publisher={Annual Reviews}
}

@Article{Leonard1980,
  author    = {Leonard, A.},
  journal   = {Journal of Computational Physics},
  title     = {Vortex methods for flow simulation},
  year      = {1980},
  issn      = {0021-9991},
  month     = oct,
  number    = {3},
  pages     = {289--335},
  volume    = {37},
  comment   = {Similar to SPH, that it is meshless},
  doi       = {10.1016/0021-9991(80)90040-6},
  publisher = {Elsevier BV},
}

@Article{Ouellette2006,
  author    = {Ouellette, Nicholas T. and Xu, Haitao and Bodenschatz, Eberhard},
  journal   = {Experiments in Fluids},
  title     = {A quantitative study of three-dimensional {L}agrangian particle tracking algorithms},
  year      = {2006},
  issn      = {1432-1114},
  month     = nov,
  number    = {2},
  pages     = {301--313},
  volume    = {40},
  comment   = {LPT fundamentals paper},
  doi       = {10.1007/s00348-005-0068-7},
  publisher = {Springer},
}

@Book{Raffel2018,
  author    = {Raffel, M. and Willert, C. E. and Scarano, F. and Kähler, C. and Wereley, S. T. and Kompenhans, J.},
  publisher = {Springer International Publishing AG},
  title     = {Particle {I}mage {V}elocimetry - {A} {P}ractical {G}uide},
  year      = {2018},
  edition   = {3},
  doi       = {10.1007/978-3-319-68852-7},
}

@Book{LeVeque2007,
  author    = {LeVeque, Randall J.},
  publisher = {Society for Industrial and Applied Mathematics},
  title     = {Finite Difference Methods for Ordinary and Partial Differential Equations: Steady-State and Time-Dependent Problems},
  year      = {2007},
  isbn      = {9780898717839},
  month     = jan,
  comment   = {Basics for finite differnces},
  doi       = {10.1137/1.9780898717839},
}

@book{Stein1999,
  title={Interpolation of Spatial Data: Some Theory for Kriging},
  author={Stein, Michael L.},
  year={1999},
  publisher={Springer},
  address={New York}
}

@Article{Candes2009,
  author    = {Candès, Emmanuel J. and Recht, Benjamin},
  journal   = {Foundations of Computational Mathematics},
  title     = {Exact {M}atrix {C}ompletion via {C}onvex {O}ptimization},
  year      = {2009},
  issn      = {1615-3383},
  month     = apr,
  number    = {6},
  pages     = {717--772},
  volume    = {9},
  doi       = {10.1007/s10208-009-9045-5},
  publisher = {Springer Science and Business Media LLC},
}

@Article{Donoho2006,
  author    = {Donoho, D.L.},
  journal   = {IEEE Transactions on Information Theory},
  title     = {Compressed sensing},
  year      = {2006},
  issn      = {0018-9448},
  month     = apr,
  number    = {4},
  pages     = {1289--1306},
  volume    = {52},
  comment   = {The compressed sensing bible/first example},
  doi       = {10.1109/tit.2006.871582},
  publisher = {Institute of Electrical and Electronics Engineers (IEEE)},
}

@article{Matulis2024,
  title={Thermal field reconstruction and compressive sensing using proper orthogonal decomposition},
  author={Matulis, John and Bindra, Hitesh},
  journal={Frontiers in Energy Research},
  volume={12},
  pages={1336540},
  year={2024},
  doi={10.3389/fenrg.2024.1336540}
}

@article{Xiong2021,
  title={Accelerating the Bayesian inference of inverse problems by using data-driven compressive sensing method based on proper orthogonal decomposition},
  author={Xiong, Meixin and Chen, Liuhong and Ming, Ju and Shin, Jaemin},
  journal={Electronic Research Archive},
  volume={29},
  number={5},
  pages={3383--3403},
  year={2021},
  doi={10.3934/era.2021044}
}

@Book{Golub2013,
  author    = {Golub, Gene H. and Van Loan, Charles F.},
  publisher = {The Johns Hopkins University Press},
  title     = {Matrix computations},
  year      = {2013},
  address   = {Baltimore},
  edition   = {Fourth edition},
  isbn      = {9781421407944},
  note      = {Literaturangaben und Index},
  series    = {Johns Hopkins studies in the mathematical sciences},
  comment   = {Very nice book for numerical methods with matrices
Also contains a nice citation on the DFT},
  pagetotal = {756},
  ppn_gvk   = {735104220},
}

@article{Fathi2018,
	title = {Dynamic {Denoising} and {Gappy} {Data} {Reconstruction} {Based} on {Dynamic} {Mode} {Decomposition} and {Discrete} {Cosine} {Transform}},
	volume = {8},
	doi = {10.3390/app8091515},
	number = {9},
	journal = {Applied Sciences},
	author = {Fathi, Mojtaba F. and Bakhshinejad, Ali and Baghaie, Ahmadreza and D’Souza, Roshan M.},
	year = {2018},
}

@article{Menon2020,
	title = {Dynamic mode decomposition based analysis of flow over a sinusoidally pitching airfoil},
	volume = {94},
	doi = {10.1016/j.jfluidstructs.2020.102886},
	journal = {Journal of Fluids and Structures},
	author = {Menon, Karthik and Mittal, Rajat},
	month = apr,
	year = {2020},
	pages = {102886},
}

@Article{Tirelli2025a,
  author    = {Tirelli, I. and Mendez, M. A. and Ianiro, A. and Discetti, S.},
  journal   = {Proceedings of the Royal Society A: Mathematical, Physical and Engineering Sciences},
  title     = {A meshless method to compute the proper orthogonal decomposition and its variants from scattered data},
  year      = {2025},
  issn      = {1471-2946},
  month     = may,
  number    = {2313},
  volume    = {481},
  comment   = {Meshless POD paper from Iacopo, just for reference},
  doi       = {10.1098/rspa.2024.0526},
  publisher = {The Royal Society},
}

@Article{Li2022,
  author   = {Binghua Li and Jesús Garicano-Mena and Eusebio Valero},
  journal  = {Journal of Computational Physics},
  title    = {A dynamic mode decomposition technique for the analysis of non–uniformly sampled flow data},
  year     = {2022},
  issn     = {0021-9991},
  pages    = {111495},
  volume   = {468},
  doi      = {10.1016/j.jcp.2022.111495},
}

@Article{Springel2010,
  author    = {Springel, Volker},
  journal   = {Annual Review of Astronomy and Astrophysics},
  title     = {Smoothed {P}article {H}ydrodynamics in {A}strophysics},
  year      = {2010},
  issn      = {1545-4282},
  number    = {Volume 48, 2010},
  pages     = {391-430},
  volume    = {48},
  comment   = {Overview of SPH in Astrophysics.
Mention binning as one of the postprocessing methods required because of the noise typical to SPH},
  doi       = {10.1146/annurev-astro-081309-130914},
  publisher = {Annual Reviews},
  type      = {Journal Article},
}

@Article{Brunton2015,
  author    = {Brunton, Steven L. and Proctor, Joshua L. and Tu, Jonathan H. and Kutz, J. Nathan},
  journal   = {Journal of Computational Dynamics},
  title     = {Compressed sensing and dynamic mode decomposition},
  year      = {2015},
  issn      = {2158-2505},
  number    = {2},
  pages     = {165--191},
  volume    = {2},
  doi       = {10.3934/jcd.2015002},
  publisher = {American Institute of Mathematical Sciences (AIMS)},
}

@Article{Taira2017,
  author  = {Taira, Kunihiko and Brunton, Steven L. and Dawson, Scott T. M. and Rowley, Clarence W. and Colonius, Tim and McKeon, Beverley J. and Schmidt, Oliver T. and Gordeyev, Stanislav and Theofilis, Vassilios and Ukeiley, Lawrence S.},
  journal = {AIAA Journal},
  title   = {Modal {A}nalysis of {F}luid {F}lows: {A}n {O}verview},
  year    = {2017},
  number  = {12},
  pages   = {4013-4041},
  volume  = {55},
  comment = {Another nice overview for POD applications},
  doi     = {10.2514/1.J056060},
}

@Article{Wu2023,
  author  = {Wu, Xiaolin and Saaid, Hicham and Voorneveld, Jason and Claessens, Tom and Westenberg, Jos and Jong, Nico and Bosch, Johan G and Kenjeres, Sasa},
  journal = {Cardiovascular Engineering and Technology},
  title   = {4D Flow Patterns and Relative Pressure Distribution in a Left Ventricle Model by Shake-the-Box and Proper Orthogonal Decomposition Analysis},
  year    = {2023},
  month   = {10},
  volume  = {14},
  doi     = {10.1007/s13239-023-00684-0},
}

@Article{Marshall2023,
  author  = {Marshall, C. and Dorrell, R. and Keevil, G. and Peakall, J. and Tobias, Steven},
  journal = {Experiments in Fluids},
  title   = {On the role of transverse motion in pseudo-steady gravity currents},
  year    = {2023},
  number  = {3},
  volume  = {64},
  doi     = {10.1007/s00348-023-03599-7},
}

@Article{Schobesberger2021,
  author  = {Schobesberger, Johannes and Worf, Dominik and Lichtneger, Petr and Yücesan, Sencer and Hauer, Christoph and Habersack, Helmut and Sindelar, Christine},
  journal = {Journal of Hydraulic Research},
  title   = {Role of low-order proper orthogonal decomposition modes and large-scale coherent structures on sediment particle entrainment},
  year    = {2021},
  month   = {04},
  volume  = {60},
  doi     = {10.1080/00221686.2020.1869604},
}

@Article{Smith2023,
  author  = {Smith, E. and Variansyah, I. and McClarren, R.},
  journal = {Nuclear Science and Engineering},
  title   = {Variable {D}ynamic {M}ode {D}ecomposition for {E}stimating {T}ime {E}igenvalues in {N}uclear {S}ystems},
  year    = {2023},
  number  = {8},
  pages   = {1769--1778},
  volume  = {197},
  doi     = {10.1080/00295639.2022.2142025},
}

@Article{Schroeder2023,
  author     = {Schröder, Andreas and Schanz, Daniel},
  journal    = {Annual Review of Fluid Mechanics},
  title      = {{3D} {L}agrangian {P}article {T}racking in {F}luid {M}echanics},
  year       = {2023},
  pages      = {511-40},
  volume     = {55},
  doi        = {10.1146/annurev-fluid-031822-041721},
  readstatus = {read},
}

@Article{Schanz2016,
  author     = {Schanz, Daniel and Gesemann, Sebastian and Schröder, Andreas},
  journal    = {Experiments in Fluids},
  title      = {{Shake}-{The}-{Box}: {Lagrangian} particle tracking at high particle image densities},
  year       = {2016},
  month      = {04},
  number     = {70},
  volume     = {57},
  doi        = {10.1007/s00348-016-2157-1},
  readstatus = {read},
}

@InProceedings{Gesemann2016,
  author     = {Sebastian Gesemann and Florian Huhn and Daniel Schanz and Andreas Schr{\"o}der},
  booktitle  = {18th International Symposium on Applications of Laser Techniques to Fluid Mechanics},
  title      = {From {Noisy} {Particle} {Tracks} to {Velocity}, {Acceleration} and {Pressure} {Fields} using {B-splines} and {Penalties}},
  year       = {2016},
  readstatus = {read},
  pages      = {1-17},
}

@Article{Jeon2022,
  author     = {Jeon, Young and Müller, Markus and Michaelis, Dirk},
  journal    = {Experiments in Fluids},
  title      = {Fine scale reconstruction {(VIC\#)} by implementing additional constraints and coarse-grid approximation into {VIC+}},
  year       = {2022},
  month      = {04},
  number     = {70},
  volume     = {63},
  doi        = {10.1007/s00348-022-03422-9},
  readstatus = {read},
}

@Article{Sperotto2022a,
  author     = {Sperotto, Pietro and Pieraccini, Sandra and Mendez, M. A.},
  journal    = {Measurement Science and Technology},
  title      = {A {M}eshless {M}ethod to {C}ompute {P}ressure {F}ields from {I}mage {V}elocimetry},
  year       = {2022},
  number     = {9},
  pages      = {094005},
  volume     = {33},
  doi        = {10.1088/1361-6501/ac70a9},
  readstatus = {read},
}

@Article{Sperotto2024b,
  author  = {Pietro Sperotto and Bo Watz and David Hess},
  journal = {Measurement Science and Technology},
  title   = {Meshless track assimilation ({MTA}) of {3D} {PTV} data},
  year    = {2024},
  pages   = {086005},
  volume  = {35},
  comment = {Used C6 RBFs as basis functions.
RBFs are also placed on a regular grid with a certain support width of the basis
Describes FlowFit as interpolating onto a regular grid.
Constraint points X_DA are kept constant over all timesteps and usually as large as the number of bases},
  doi     = {10.1088/1361-6501/ad3f36},
}

@Article{Sciacchitano2025,
  author  = {A. Sciacchitano and B. Leclaire and A. Schröder},
  journal = {Experiments in Fluids},
  title   = {On the accuracy of data assimilation algorithms for dense flow field reconstructions},
  year    = {2025},
  number  = {42},
  volume  = {66},
  doi     = {10.1007/s00348-025-03969-3},
}

@Article{Sperotto2024a,
  author  = {P. Sperotto and M. Ratz and M. A. Mendez},
  journal = {Journal of Open Source Software},
  title   = {{SPICY}: a {Python} toolbox for meshless assimilation from image velocimetry using radial basis functions},
  year    = {2024},
  number  = {93},
  volume  = {9},
  doi     = {10.21105/joss.05749},
}

@Book{Hastie2009,
  author    = {Hastie, Trevor and Tibshirani, Robert and Friedman, Jerome},
  publisher = {Springer New York},
  title     = {The Elements of Statistical Learning},
  year      = {2009},
  isbn      = {9780387848587},
  doi       = {10.1007/978-0-387-84858-7},
  issn      = {2197-568X},
  journal   = {Springer Series in Statistics},
}

@Book{Bishop2011,
  author    = {Christopher M. Bishop},
  publisher = {Springer},
  title     = {Pattern Recognition and Machine Learning},
  year      = {2011},
}

@Article{Li2024,
  author  = {Lanyu Li and Zhao Pan},
  journal = {Experiments in Fluids},
  title   = {Three-dimensional time-resolved {L}agrangian flow field reconstruction based on constrained least squares and stable radial basis function},
  year    = {2024},
  pages   = {57},
  volume  = {65},
  comment = {Give references to the application of LPT.
RBF kernel based QR oversampling regression
Use the divergence as a temporal corrector for 3D assimilation in each timestep},
  doi     = {10.1007/s00348-024-03788-y},
}

@Article{Aguei1987,
  author     = {Agüí, Juan C. and Jiménez, J.},
  journal    = {Journal of Fluid Mechanics},
  title      = {On the performance of particle tracking},
  year       = {1987},
  pages      = {447–468},
  volume     = {185},
  comment    = {Adaptive Gaussian Weighting, original publication},
  doi        = {10.1017/S0022112087003252},
  publisher  = {Cambridge University Press},
  readstatus = {read},
}

@Article{Tan2020,
  author  = {Shiyong Tan and Ashwanth Salibindla and Ashik Ullah Mohammad Masuk and Rui Ni},
  journal = {Experiments in Fluids},
  title   = {Introducing {OpenLPT}: new method of removing ghost particles and high‑concentration particle shadow tracking},
  year    = {2020},
  number  = {47},
  volume  = {61},
  doi     = {10.1007/s00348-019-2875-2},
}

@Article{Tirelli2023b,
  author     = {Iacopo Tirelli and Andrea Ianiro and Stefano Discetti},
  journal    = {Experimental Thermal and Fluid Science},
  title      = {A simple trick to improve the accuracy of {PIV/PTV} data},
  year       = {2023},
  issn       = {0894-1777},
  pages      = {110872},
  volume     = {145},
  comment    = {Better computation to first obtain the mean and subtract it in every point.},
  doi        = {10.1016/j.expthermflusci.2023.110872},
  readstatus = {read},
}

@Article{Ratz2024,
  author  = {Manuel Ratz and Miguel Alfonso Mendez},
  journal = {Experiments in Fluids},
  title   = {A meshless and binless approach to compute statistics in {3D} ensemble {PTV}},
  year    = {2024},
  number  = {142},
  volume  = {65},
  doi     = {10.1007/s00348-024-03878-x},
}

@misc{Rigutto2025,
    author = {Damien Rigutto and Manuel Ratz and Miguel Alfonso Mendez},
    howpublished = {Submitted to: Experiments in Fluids},
    title = {A meshless data-tailored approach to compute statistics from scattered data with adaptive radial basis functions},
    year = {2025},
}

@Article{Oliver1990,
  author    = {M. A. Oliver and R. Webster},
  journal   = {International journal of geographical information systems},
  title     = {Kriging: a method of interpolation for geographical information systems},
  year      = {1990},
  number    = {3},
  pages     = {313--332},
  volume    = {4},
  comment   = {Gold-standard paper for kriging interpolation},
  doi       = {10.1080/02693799008941549},
  publisher = {Taylor \& Francis},
}

@Article{CortinaFernandez2021,
  author     = {J. Cortina-Fernández and C. {Sanmiguel Vila} and A. Ianiro and S. Discetti},
  journal    = {Experimental Thermal and Fluid Science},
  title      = {From sparse data to high-resolution fields: ensemble particle modes as a basis for high-resolution flow characterization},
  year       = {2021},
  issn       = {0894-1777},
  pages      = {110178},
  volume     = {120},
  comment    = {Gappy POD together with EPTV to obtain high-resolution instantaneous fields
Use low resolution POD from PIV and then project the interpolated PTV data onto this grid},
  doi        = {10.1016/j.expthermflusci.2020.110178},
  readstatus = {skimmed},
}

@Article{GrilleGuerra2024b,
  author   = {Grille Guerra, Adrian and Sciacchitano, Andrea and Scarano, Fulvio},
  journal  = {Physics of Fluids},
  title    = {Iterative modal reconstruction for sparse particle tracking data},
  year     = {2024},
  issn     = {1070-6631},
  month    = {07},
  number   = {7},
  pages    = {075107},
  volume   = {36},
  doi      = {10.1063/5.0209527},
}

@Article{Gunes2006,
  author  = {Gunes, Hasan and Sirisup, Sirod and Karniadakis, George},
  journal = {Journal of Computational Physics},
  title   = {Gappy data: To Krig or not to Krig?},
  year    = {2006},
  month   = {02},
  pages   = {358-382},
  volume  = {212},
  comment = {One of the earliest example of efficient gappy POD},
  doi     = {10.1016/j.jcp.2005.06.023},
}

@article{Nekkanti2023,
	title = {Gappy spectral proper orthogonal decomposition},
	volume = {478},
	doi = {10.1016/j.jcp.2023.111950},
	journal = {Journal of Computational Physics},
	author = {Nekkanti, Akhil and Schmidt, Oliver T.},
	year = {2023},
	pages = {111950--111950},
}

@Article{Everson1995,
  author  = {Everson, Richard and Sirovich, Lawrence},
  journal = {Journal of the Optical Society of America A},
  title   = {Karhunen–{L}oève procedure for gappy data},
  year    = {1995},
  number  = {8},
  pages   = {1657--1664},
  volume  = {12},
  comment = {Gappy KL transform (POD for},
  doi     = {10.1364/JOSAA.12.001657},
}

@Article{Mendez2017,
  author     = {Mendez, Miguel and Raiola, Marco and Masullo, Alessandro and Discetti, Stefano and Ianiro, Andrea and Theunissen, R. and Buchlin, Jean-Marie},
  journal    = {Experimental Thermal and Fluid Science},
  title      = {{POD}-based {B}ackground {R}emoval for {P}article {I}mage {V}elocimetry},
  year       = {2017},
  month      = {01},
  pages      = {181--192},
  volume     = {80},
  comment    = {POD background removal paper from Miguel},
  doi        = {10.1016/j.expthermflusci.2016.08.021},
  readstatus = {read},
}

@Article{Kirby1990,
  author   = {Kirby, M. and Sirovich, L.},
  journal  = {IEEE Tr`ansactions on Pattern Analysis and Machine Intelligence},
  title    = {Application of the Karhunen-Loeve procedure for the characterization of human faces},
  year     = {1990},
  number   = {1},
  pages    = {103-108},
  volume   = {12},
  comment  = {Application of POD to human face characterization},
  doi      = {10.1109/34.41390},
}

@Article{Fassois2007,
  author   = {Fassois, Spilios D and Sakellariou, John S},
  journal  = {Philosophical Transactions of the Royal Society A: Mathematical, Physical and Engineering Sciences},
  title    = {Time-series methods for fault detection and identification in vibrating structures},
  year     = {2007},
  number   = {1851},
  pages    = {411-448},
  volume   = {365},
  comment  = {Application of POD to structural analysis},
  doi      = {10.1098/rsta.2006.1929},
}

@Article{Troje2002,
  author   = {Troje, Nikolaus F.},
  journal  = {Journal of Vision},
  title    = {Decomposing biological motion: A framework for analysis and synthesis of human gait patterns},
  year     = {2002},
  issn     = {1534-7362},
  month    = {09},
  number   = {5},
  pages    = {371--387},
  volume   = {2},
  comment  = {Application of POD in biomechanics},
  doi      = {10.1167/2.5.2},
}

@Article{Zhang2000,
  author  = {Zhang, X. and Song, K. and Lu, M.},
  journal = {Computational Mechanics},
  title   = {Meshless methods based on collocation with radial basis functions},
  year    = {2000},
  pages   = {333--343},
  volume  = {26},
  doi     = {10.1007/s004660000181},
}

@Book{Chen2014,
  author    = {Wen Chen and Zhuo-Jia Fu and C. S. Chen},
  publisher = {Springer Berlin, Heidelberg},
  title     = {Recent {A}dvances in {R}adial {B}asis {F}unction {C}ollocation {M}ethods},
  year      = {2014},
  edition   = {1},
  comment   = {Describe the polynomial embedding that was used introduced by Kansa 1990. They also mention that it is no longer necessary.
Good reference for why our world is meshless},
  doi       = {10.1007/978-3-642-39572-7},
}

@Article{Fornberg2015,
  author    = {Fornberg, Bengt and Flyer, Natasha},
  journal   = {Acta Numerica},
  title     = {Solving {PDE}s with radial basis functions},
  year      = {2015},
  pages     = {215–258},
  volume    = {24},
  comment   = {Good overview over the basics of RBFs},
  doi       = {10.1017/S0962492914000130},
  publisher = {Cambridge University Press},
}

@Book{Fasshauer2007,
  author    = {Fasshauer, Gregory E.},
  publisher = {World Scientific},
  title     = {Meshfree approximation methods with {MATLAB}},
  year      = {2007},
  isbn      = {978-981-270-634-8},
  volume    = {6},
}

@Misc{Fong2011,
  author       = {David Chin-Lung Fong and Michael Saunders},
  howpublished = {arxiv},
  title        = {{LSMR}: An iterative algorithm for sparse least-squares problems},
  year         = {2011},
  comment      = {LSMR arxiv preprint},
  doi          = {10.48550/arXiv.1006.0758},
}

@article{ramsay2005principal,
  title={Principal components analysis for functional data},
  author={Ramsay, JO and Silverman, BW},
  journal={Functional data analysis},
  pages={147--172},
  year={2005},
  publisher={Springer},
doi = {10.1007/0-387-22751-2_8}

}

@article{Wang2016,
  title   = {A Survey of Functional Principal Component Analysis},
  author  = {Jane-Ling Wang and Jeng-Min Chiou and Hans-Georg Müller},
  journal = {The American Statistician},
  year    = {2016},
  volume  = {70},
  number  = {2},
  pages   = {127--137},
  doi     = {10.1080/00031305.2016.1141704}
}

@article{Hall2006,
  title   = {On Properties of Functional Principal Components Analysis},
  author  = {Peter Hall and Joel L. Horowitz},
  journal = {Journal of the Royal Statistical Society: Series B (Statistical Methodology)},
  year    = {2006},
  volume  = {68},
  number  = {1},
  pages   = {109--126},
  doi     = {10.1111/j.1467-9868.2005.00535.x}
}

@Misc{Crameri2023,
  author    = {Crameri, Fabio},
  title     = {Scientific colour maps},
  year      = {2023},
  copyright = {MIT License},
  doi       = {10.5281/ZENODO.8409685},
  language  = {en},
  publisher = {Zenodo},
}

@Article{Crameri2020,
  author    = {Crameri, Fabio and Shephard, Grace E. and Heron, Philip J.},
  journal   = {Nature Communications},
  title     = {The misuse of colour in science communication},
  year      = {2020},
  issn      = {2041-1723},
  month     = oct,
  number    = {1},
  volume    = {11},
  comment   = {My favorite reading about colormaps. Proposes perceptually uniform colormaps for all sorts of applications.
Also an excellent guide on how to choose the colormaps},
  doi       = {10.1038/s41467-020-19160-7},
  publisher = {Springer Science and Business Media LLC},
}

@Book{Holmes2012,
  author    = {Holmes, Philip and Lumley, John L. and Berkooz, Gahl and Rowley, Clarence W.},
  publisher = {Cambridge University Press},
  title     = {Turbulence, {C}oherent {S}tructures, {D}ynamical {S}ystems and {S}ymmetry},
  year      = {2012},
  isbn      = {9781107008250},
  month     = feb,
  doi       = {10.1017/cbo9780511919701},
}

@Article{Procacci2022,
  author    = {Procacci, A. and Kamal, M. M. and Mendez, M. A. and Hochgreb, S. and Coussement, A. and Parente, A.},
  journal   = {Physics of Fluids},
  title     = {Multi-scale proper orthogonal decomposition analysis of instabilities in swirled and stratified flames},
  year      = {2022},
  issn      = {1089-7666},
  month     = dec,
  number    = {12},
  volume    = {34},
  doi       = {10.1063/5.0127956},
  publisher = {AIP Publishing},
}


\appendix

\section{Nomenclature} \label{sec:appA}

We use lowercase letters for scalar continuous quantities, e.g. $t\in\mathbb{R}$. Bold lowercase letters are used for vector valued continuous quantities, e.g. $\boldsymbol{x}\in\mathbb{R}^{d}$. 
We use the roman bold symbol to denote vectors collecting samples of a scalar quantity, e.g. $\mathbf{t}$ or $\mathbf{c}$, and the subscript $\mathbf{t}_k$ or $\mathbf{c}_m$ denote the k-th or the m-th entry of these vectors. Similarly, a collection of vector valued quantities is indicated with a roman bold capital symbol, hence $\mathbf{X}\in\mathbb{R}^{d\times n}$ is a collection of $n$ vectors of size $d$. The subscript on a matrix is generally used to denote its column, hence $\mathbf{X}_m\in\mathbb{R}^{d}$ is the m-th column of $\mathbf{X}$, thus the m-th vector valued sample. However, sometimes a subscript is used to distinguish variables: for example $\mathbf{X}_p$ is the set of training points while $\mathbf{X}_c$ is the set of collocation points. In this case, the sampling is always used as a second script, e.g. $\mathbf{X}_{c,m}$ is the m-th collocation point. Confusion in the usage of subscripts for different purposes can be easily avoided by noticing that only seven letters are used for indices: these are $i,j,r,s,l,m,n$. Any other subscript (such as p or c) is thus not an index.

We use the square brackets to denote vectors or matrices created from collections of scalars or vectors. Therefore, $\mathbf{t}=[\mathbf{t}_1, \mathbf{t}_2, \cdots , \mathbf{t}_{n_t} ]= \{\mathbf{t}_i\}^{n_t}_{i=1}$ is the vector collecting the times of all spaced samples in time while $\mathbf{X}=[\mathbf{X}_1, \mathbf{X}_2 \dots \mathbf{X}_{n_p}]=\{\mathbf{X}_n\}^{n_p}_{n=1}$ is a matrix collecting $n_p$ location in space $\mathbf{X}_n$. In either case, we use the curly brackets $\{ \}$ for sequences of scalars or vectors.

Equivalently, when stress is placed on the specific scalars forming a vector or vectors forming a matrix, a superscript with parenthesis is used. For example $\hat{\mathbf{W}}_{\mathcal{W},u}=[\hat{\mathbf{w}}_u^{(1)},\hat{\mathbf{w}}_u^{(2)}\cdots \hat{\mathbf{w}}_u^{(n_B)}]$ is the matrix built by stacking the DFT of the weights $\mathbf{w}_u$, denoted as $\hat{\mathbf{w}}_u$, at each of the $n_B$ blocks of the Welch averaging procedure.

Concerning the indices of entries in vectors and matrices, we adopt a Python-like notation. Specifically, $\boldsymbol{A}[i,j]$ denotes the entry in the $i$-th row and $j$-th column of the matrix $\boldsymbol{A}$, while $\mathbf{v}[i]$ refers to the $i$-th entry of a vector $\mathbf{v}$. Accordingly, $\boldsymbol{A}[i,j] = \boldsymbol{A}_j[i]$ are two equivalent ways of referencing the same entry. We retain both notations: the former emphasizes how certain matrices are constructed, while the latter highlights the role of certain column vectors. Moreover, since vectors and matrix columns typically collect samples in time, space, or frequency domains, index and function notations are both relevant and used. Thus the notation $ \hat{\boldsymbol{u}}(\boldsymbol{x}, \mathbf{f}_l)$ emphasizes that $\hat{\boldsymbol{u}}$ is a continuous function in $\boldsymbol{x}$ but a discrete function in $\mathbf{f}_l$ and the column $\mathbf{W}_{u,m}=\mathbf{w}_u (\mathbf{t}_m)$ collects the RBF weight vector at the m-th time step.

\renewcommand{\nomname}{}


\nomenclature[S$^$]{$\hat{q}$}{discrete Fourier transform}
\nomenclature[S$^\ast$]{${q}^\ast$}{complex conjugate or hermitian transpose of $q$}
\nomenclature[S$^\tilde$]{${\tilde{\mathbf{q}}}$}{truncated to $\tilde{R}$ POD modes}
\nomenclature[S$^\T$]{${\mathbf{q}^\T}$}{transpose}
\nomenclature[S$_{\mathcal{P}}$]{$q_{\mathcal{P}}$}{derived from a proper orthogonal decomposition}

\nomenclature[N$St$]{St}{Strouhal}

\nomenclature[MT1]{$\langle p,q \rangle_T$}{inner product in time}
\nomenclature[MT2]{$\langle p,q \rangle_{T,d}$}{inner product in time from discrete samples}
\nomenclature[MS1]{$\langle p,q \rangle_\Omega$}{inner product in space}
\nomenclature[MS2]{$\langle p,q \rangle_{\Omega,a}$}{inner product in space from discrete samples}
\nomenclature[MS3]{$\langle p,q \rangle_{\Omega,d}$}{inner product in space from an analytical expression}

\nomenclature[AmPOD]{mPOD}{multi-scale proper orthogonal decomposition}
\nomenclature[ASPOD]{SPOD}{spectral proper orthogonal decomposition}
\nomenclature[ADMD]{DMD}{dynamic mode decomposition}
\nomenclature[ADFT]{DFT}{discrete fourier transform}
\nomenclature[ARBF]{RBF}{radial basis fuction}
\nomenclature[AFPCA]{FPCA}{functional principal component analysis}
\nomenclature[APIV]{PIV}{particle image velocimetry}
\nomenclature[APSD]{PSD}{power spectral density}
\nomenclature[ALES]{LES}{large eddy simulation}
	
\printnomenclature

\end{document}